\documentclass{iopart}
\usepackage{graphicx}

\newcommand{\mb}[1]{\mbox{\boldmath $#1$}}
\def\be{\begin{equation}}
\def\ee{\end{equation}}
\def\beq{\begin{eqnarray}}
\def\eeq{\end{eqnarray}}
\def\cM{{\cal M}}
\def\cN{{\cal N}}
\def\cL{{\cal L}}

\def\RR{{\rm I\!R}}
\def\NN{{\rm I\!N}}
\def\nn{\nonumber}
\def\lo{{\!{}_{\lambda=\Omega=0}}}

\begin{document}
\jl{6}

\title[Two--parameter non--linear spacetime
perturbations]{Two--parameter non--linear spacetime perturbations: gauge
transformations and gauge invariance}

\author{
Marco Bruni\ddag\, Leonardo Gualtieri\S\ and Carlos F. Sopuerta\ddag}
\address{\ddag\ Institute of Cosmology and Gravitation, Mercantile
House, University of Portsmouth, Portsmouth PO1 2EG, Britain}
\address{\S\ Dipartimento di Fisica ''G. Marconi'', Universit\`a di Roma
''La Sapienza'' and Sezione INFN ROMA 1, piazzale Aldo Moro 2, I--00185 Roma,
Italy}

\date{\today}

%
%

\begin{abstract}
An implicit fundamental assumption in relativistic perturbation theory is
that there exists a parametric family of spacetimes that can be Taylor
expanded around a background.
The choice of the latter is crucial to obtain a manageable theory, so
that it is sometime convenient to construct a perturbative formalism
based on two (or more) parameters.  The study of perturbations of
rotating stars is a good example: in this case one can
treat the stationary axisymmetric star using a slow rotation approximation
(expansion in the angular velocity $\Omega$), so that the background is
spherical.
Generic perturbations of the rotating star (say parametrized
by  $\lambda$) are then built on top of the axisymmetric
perturbations in $\Omega$.  Clearly, any interesting physics requires
non--linear perturbations, as at least terms $\lambda\Omega$ need to
be considered.
In this paper we analyse the gauge dependence of non--linear
perturbations depending on two parameters, derive explicit higher
order gauge transformation rules, and define
gauge invariance. The formalism is completely
general and can be used in different applications of general
relativity or any other spacetime theory.
\end{abstract}
\maketitle


%
%

\section{Introduction}

An implicit fundamental assumption in relativistic perturbation
theory is that there exists a parametric family of spacetimes such that
the perturbative formalism is built as a Taylor expansion of this
family around a background.  The perturbations are then
defined as the derivative terms of this series, evaluated on this
background~\cite{wald}. In most cases of interest one deals with an expansion
in a single parameter, which can either be a formal one, as in
cosmology~\cite{BMMS,MMB,BMT} or in the study of quasi--normal modes of stars
and black holes~\cite{chandrabook,kokkomodes},
or can have a specific physical meaning, as in the study of binary
black hole mergers via the close limit approximation~\cite{price,gleisclose},
or in the study of quasi--normal mode excitation by a physical source (see
\cite{kokkomodes,my123} and references therein). Typically
 the perturbative expansion stops at the first order, but recent
interesting developments deal with second order
perturbations~\cite{BMMS,MMB,BMT,cl,gleiser}.

In some physical applications it may be instead convenient to
construct a perturbative formalism based on two (or more) parameters,
because the choice of background is crucial in having a manageable
theory.  The study of perturbations of stationary axisymmetric rotating
stars (see \cite{kojimarev,laf,rsk} and references therein) is a good example.
In this case, an analytic stationary axisymmetric solution
is not known, at least for reasonably interesting equations of state.
A common procedure is to treat axisymmetric stars using the
so--called slow rotation approximation, so that the background is a
star with spherical symmetry~\cite{hartle,hartlethorne}. In this approach
the first order in $\Omega$ discloses frame dragging effects, with the
star actually
remaining spherical; $\Omega^2$ terms carry the effects of rotation on
the fluid. This is intuitive from a Newtonian point of view,
as rotational kinetic energy goes like $\Omega^{2}$.
This approximation is
valid for angular velocities $\Omega$ much smaller than the mass
shedding limit $\Omega_K\equiv\sqrt{M/R_{star}^3}$, with typical
values for neutron stars $\Omega_K\sim 10^3Hz$. Therefore the
slow rotation approximation, despite the name, can still be valid for large
angular velocities. In practice,
the perturbative approach up to $\Omega^2$ is accurate for
most astrophysical situations, with the exception of newly born
neutron stars (see~\cite{STER} and references therein).

Given that the differential operators
appearing in the perturbative treatment of a problem are those defined
on the background, the theory is considerably simplified when the
latter is spherical.  Generic time dependent perturbations of the
rotating star (parametrized by a dummy parameter $\lambda$ and
describing oscillations)
are then built on top of the stationary axisymmetric perturbations
in $\Omega$.  Clearly, in this approach any interesting physics
requires non--linear perturbations, as at least terms of order
$\lambda\Omega$ need to be considered.  A similar approach could be
used to study perturbations of the slowly rotating collapse, even if in
specific cases~\cite{CPM,seidel,gundlach}
the perturbative expansion depends by one parameter only.

Classical studies in the literature have not analysed in full the
gauge dependence and gauge invariance of the non-linear perturbation
theory.  For example, in~\cite{CPM} the second order perturbations  are
treated in a gauge invariant fashion on top of the first order perturbation
in a given specific gauge.
The perturbation variables used are therefore non gauge invariant
under a complete second order gauge transformation~\cite{BMMS,FW},
but only invariant under  ``first order transformations acting at
second order''~\cite{CPM}.
While this may
be perfectly satisfactory from the point of view of obtaining
physical results, one may wish to convert results in a given gauge to a
different one~\cite{BMMS,MMB}, to compare results obtained in two different
gauges, or to construct a fully gauge invariant
formalism.  To this end one needs to know the gauge transformation
rules and the rules for gauge invariance, either {\it up to} order
$n$~\cite{BMMS} or {\it at} order $n$ {\it only}, as in~\cite{CPM}.
The situation is going to be more complicated in the case of two
parameters, as we shall see.\footnote{The concept of perturbation theory with
more than one parameter has already been introduced, for first-order
perturbations, in~\cite{MAIA},
where the standard definition of spacetime perturbations~\cite{stewart} is
extended
by using a (4+n)-dimensional flat space in which space-times are embedded.  The
main aim
of these works is to re-examine the gauge invariance of the metric.
}

In this paper we keep in mind the above practical examples, but we do
not make any specific assumption on the background spacetime and the
two--parameter family it belongs to. As in \cite{BMMS,SB,BS},
we do not even need to
assume that the background is a solution of Einstein's field equations:
the formalism is completely general  and can be applied to any
spacetime theory.  We analyse the gauge dependence
of perturbations in the case when they depend on two parameters, $\lambda$
and $\Omega$, derive explicit gauge transformation rules up to fourth order,
i.e.\ including any term $\lambda^k\Omega^{k'}$ with $k+k'\leq 4$,
and define gauge invariance.
This choice of keeping fixed the total perturbative order is due to
the generality of our approach.  In practical applications one would
be guided by the physical characteristics of the problem in deciding
where to truncate the perturbative expansion.  For example, in the
case of a rotating star one could consider first order oscillations,
parametrized by $\lambda$, on top of a stationary axisymmetric
background described up to $\Omega^2$, neglecting therefore
$\lambda^2\Omega$ terms.  Or instead, one could decide that
$\lambda^2\Omega$ terms are more interesting than the
$\lambda\Omega^2$ ones in certain cases.

From a practical point of view, our aim is to derive the effects
of gauge transformations on tensor fields $T$ up to order $k+k'=4$.
It is indeed reasonable to assume that in a practical example like
that of rotating stars, at most one will want to consider second order
oscillations~$\sim\lambda^2$ on top of a slowly rotating background
described up to $O(\Omega^2)$, in order to take into account
large oscillations and fluid deformations due to rotation.

We will show that the coordinate form of a
two--parameter gauge transformation can be represented by:
\beq
\fl {\tilde x}^{\mu} & = & x^{\mu}+\lambda\xi^{\mu}_{(1,0)}+
\Omega\xi^{\mu}_{(0,1)} \nn \\
\fl & &
+\frac{\lambda^2}{2}\left(\xi^{\mu}_{(2,0)}+\xi^{\nu}_{(1,0)}
\xi^{\mu}_{(1,0),\nu}\right)
+\frac{\Omega^2}{2}\left(\xi^{\mu}_{(0,2)}+\xi^{\nu}_{(0,1)}
\xi^{\mu}_{(0,1),\nu}\right) \nn \\
\fl & &
+\lambda\Omega\left(\xi^{\mu}_{(1,1)}+\epsilon_0\xi^{\nu}_{(1,0)}
\xi^{\mu}_{(0,1),\nu}+\epsilon_1\xi^{\nu}_{(0,1)}
\xi^{\mu}_{(1,0),\nu}\right)+O^3(\lambda,\Omega)\,, \label{coord2}
\eeq
where the full expression is given in Eq.~(\ref{coordtransf}).  Here
$\xi^\mu_{(1,0)}$, $\xi^\mu_{(0,1)}$, $\xi^\mu_{(2,0)}$,
$\xi^\mu_{(1,1)}$, and $\xi^\mu_{(0,2)}$ are independent vector
fields and $(\epsilon_0,\epsilon_1)$ are any two real numbers
satisfying $\epsilon_0+\epsilon_1=1$.
Coupling terms like the $\lambda\Omega$ in (\ref{coord2}) are the expected new
features of the two--parameter case, cf.\ \cite{BMMS,SB}.
Our main results are the explicit transformation rules
for the perturbations of a tensor field $T$ and the
conditions for the gauge invariance of these perturbations.

The paper is organized as follows: in
Section~\ref{taylorexp} we develop the necessary mathematical tools,
deriving Taylor expansion formulae for two--parameter groups of
diffeomorphisms and for general two--parameter families of
diffeomorphisms.  In Section~\ref{gauges}
we set up an appropriate geometrical description of the gauge
dependence of perturbations in the specific case of two--parameter
families of spacetimes.  In Section~\ref{gaugeiandt} we apply the
tools developed in Section~\ref{taylorexp} to the framework introduced
in Section~\ref{gauges}, in order to define gauge invariance
and formulas for gauge transformations, up to fourth
order in the two--parameter perturbative expansion.
Section~\ref{conclusions} is devoted to the conclusions.
We follow the notation  used previously in
\cite{BMMS,SB,BS} for the case of one parameter perturbations.

%
%

\section{Taylor expansion of tensor fields}\label{taylorexp}

In order to consider the issues of gauge transformations and
gauge invariance in two--parameter perturbation theory we
need first to introduce some mathematical tools concerning the
two--parameter Taylor expansion of tensor fields.  Since Taylor
expansions are aimed to provide the value of a quantity at
some point in terms of its value, and the value of its derivatives,
at another point, a Taylor expansion of tensorial quantities
can only be defined through a mapping between tensors at different
points of the manifold under consideration.
In this section we consider the cases where such a mapping is
given by a two--parameter family of diffeomorphisms of $\cM$,
starting from the simplest case in which such a family constitutes
a group.

\subsection{Two--parameter groups of diffeomorphisms}
\label{groupdif}
Given a differentiable manifold $\cM$, a two--parameter group of
diffeomorphisms $\phi$ of $\cM$ can be represented as follows
\be
\begin{array}{cccc}
\phi:&\cM\times\RR^2 & \longrightarrow & \cM\times\RR^2 \\
&(p,\lambda,\Omega)&|\!\!\!\longrightarrow&
(\phi_{\lambda,\Omega}(p),\lambda,\Omega)\,.  \\
\end{array}
\ee
For the purpose of introducing two-parameter perturbation
theory, where perturbing first with respect to the parameter $\lambda$
and afterwards with respect to the parameter $\Omega$ should
be equivalent to the converse operation, we will assume that
$\phi_{\lambda,\Omega}$ is such that it satisfies the following
property\footnote{We adopt the convention that the first label in $\phi$
corresponds to the flow generated by $\eta$ and parametrized by $\lambda$,
while the second label corresponds to the flow generated by $\zeta$ and
parametrized by $\Omega$. Therefore $\phi_{\lambda,\Omega}\neq
\phi_{\Omega,\lambda}$.}
\beq
\phi_{\lambda_1,\Omega_1}\circ\phi_{\lambda_2,\Omega_2}=
\phi_{\lambda_1+\lambda_2,\Omega_1+\Omega_2}\,,~~~
\forall\;\lambda,\Omega\in\RR\,.
\eeq
By one hand, this property implies that the two-parameter
group is Abelian.  On the other hand, it allow us to
make the following useful decomposition of $\phi_{\lambda,\Omega}$
into two one--parameter groups of diffeomorphisms (flows)
that remain implicitly defined by the equalities
\be
\phi_{\lambda,\Omega}=\phi_{\lambda,0}\circ\phi_{0,\Omega}
=\phi_{0,\Omega}\circ\phi_{\lambda,0}\,. \label{philambda}
\ee
The action of the flows $\phi_{\lambda,0}$ and $\phi_{0,\Omega}$
is generated by two vector fields, $\eta$ and $\zeta$ respectively, acting
on the tangent space of $\cM\times\RR^2$.  The Lie derivatives of a generic
tensor $T$ with respect to $\eta$ and $\zeta$
are
\beq
\fl\cL_{\eta}T&: =
\lim_{\lambda\rightarrow 0}\frac{1}{\lambda}
(\phi^*_{\lambda,0}T-T)  =
&\!\left[\frac{d}{d\lambda}\phi^*_{\lambda,0}T\right]_{\lambda=0}\!\!\!
=\left[\frac{\partial}{\partial\lambda}\phi^*_{\lambda,\Omega}
T\right]_{\lo}\,,  \\
\fl\cL_{\zeta}T&:=
\lim_{\Omega\rightarrow 0}\frac{1}{\Omega}
(\phi^*_{0,\Omega}T-T) =
&\!\left[\frac{d}{d\Omega}\phi^*_{0,\Omega}T\right]_{\Omega=0}\!\!\!
=\left[\frac{\partial}{\partial\Omega}\phi^*_{\lambda,\Omega}
T\right]_{\lo}\,,
\eeq
where the superscript ${}^*$ denotes the pull--back map associated with
the corresponding diffeomorphism~\cite{BMMS}.  Because the group
is Abelian  the vector fields $\eta$ and $\zeta$ must commute
\be
[\eta,\zeta]=0\,.
\ee
The Taylor expansion of the pull--backs
$\phi^*_{\lambda,0} T,\phi^*_{0,\Omega} T$ is given by (see~\cite{BMMS})
\beq
\phi^*_{\lambda,0}T&=\sum_{k=0}^{\infty}\frac{\lambda^k}{k!}
\left[\frac{d^k}{d\lambda^k}\phi^*_{\lambda,0}T\right]_{\lambda=0}=
&\sum_{k=0}^{\infty}\frac{\lambda^k}{k!}\cL^k_{\eta}T \,, \\
\phi^*_{0,\Omega}T&=\sum_{k=0}^{\infty}\frac{\Omega^k}{k!}
\left[\frac{d^k}{d\Omega^k}\phi^*_{0,\Omega}T\right]_{\Omega=0}=
&\sum_{k=0}^{\infty}\frac{\Omega^k}{k!}
\cL^k_{\zeta}T\,.
\eeq
From this, using (\ref{philambda}), we can
derive the Taylor expansion of the two--parameter
group of pull--backs $\phi^*_{\lambda,\Omega}T$:
\be
\fl \phi^*_{\lambda,\Omega}T = \sum_{k,k'=0}^{\infty}
\frac{\lambda^k{\Omega^{k}}'}{k!\,k'!}\left[\frac{\partial^{k+k'}}
{\partial \lambda^k\partial\Omega^{k'}}\phi^*_{\lambda,\Omega}
T\right]_{\lo} = \sum_{k,k'=0}^{\infty}
\frac{\lambda^k{\Omega^k}'}{k!\,k'!}
\cL^k_{\eta}\cL^{k'}_{\zeta}T\,.\label{Taylorgroup}
\ee

\subsection{Two--parameter families of diffeomorphisms}
\label{familydiffeo}

Let us now consider the general case of a two--parameter family of
diffeomorphisms $\Phi$:
\be
\begin{array}{cccc}
\Phi:&\cM\times\RR^2 & \longrightarrow & \cM\times\RR^2\\
&(p,\lambda,\Omega)&|\!\!\!\longrightarrow&
(\Phi_{\lambda,\Omega}(p),\lambda,\Omega) \,.\\
\end{array}
\ee
In this generic case the diffeomorphisms $\Phi_{\lambda,\Omega}$ do
not form in general a group.  In particular,
\beq
\Phi_{\lambda_1,\Omega_1}\circ\Phi_{\lambda_2,\Omega_2}\neq
\Phi_{\lambda_1+\lambda_2,\Omega_1+\Omega_2}\,, \nn
\eeq
which means that we cannot decompose
$\Phi_{\lambda,\Omega}$ as in the case of a two--parameter group of
diffeomorphisms.   The Taylor expansion of the pull--back
of $\Phi_{\lambda,\Omega}$ is formally given by
\be
\Phi^*_{\lambda,\Omega} T= \sum_{k,k'=0}^{\infty}
\frac{\lambda^k{\Omega^{k}}'}{k!\,k'!}\left[\frac{\partial^{k+k'}}
{\partial \lambda^k\partial\Omega^{k'}}\Phi^*_{\lambda,\Omega}
T\right]_{\lo}\,.
\label{Taylorfamily}
\ee
Since the diffeomorphisms $\Phi_{\lambda,\Omega}$ do not form
a group we cannot write this expansion directly in terms
of Lie derivatives as in the previous case, Eq.~(\ref{Taylorgroup}).
Nevertheless, in order to study the characteristics and properties
of the gauge transformations to be derived in Section \ref{gaugeiandt},
we would like to find an alternative way of expressing the
expansion~(\ref{Taylorfamily}) in terms of suitable Lie derivatives,
in a similar way as it was done in~\cite{BMMS,BS} in the one--parameter case.
To this end,  new objects called {\em knight
diffeomorphisms} were introduced in~\cite{BMMS,BS}.  Broadly speaking, they are
a composition of one-parameter
groups of diffeomorphisms that can reproduce the action of a given family
of diffeomorphisms, in such a way that the more groups  we compose the
better the approximation is.  When the number of composed groups tends
to infinity the reproduction of the family is exact (see~\cite{BMMS,SB,BS} for a
detailed account on knight diffeomorphisms).
The main aim for using these objects was to
show~\cite{BMMS,BS} that it is possible to expand a one-parameter
family of diffeomorphism, at orden $n$, using Lie derivatives with respect
to a finite number $n$ of vector fields.  Knight diffeomorphisms constitute an
elegant formulation
of this kind of expansions.  In order to apply these ideas to
our case, we can think of translating the idea of knight diffeomorphims
from the one-parameter to the two-parameter case.  We have studied
this question and we have found  that in the two--parameter case 
there are several ways of formulating knight diffeomorphisms, and they lead to
different formal expansions
of a two-parameter family of diffeomorphisms.
Actually, we have found that some of these formulations can
be inconsistent in the sense that the differential operators that
come out from them do not satisfy the Leibnitz rule. Then,
we cannot associated to any of these operators a vector field whose
Lie derivative will describe the action of that operator.
Despite of these facts, we find that the main goal for introducing
knight diffeomorphisms can be achieved in the two-parameter case.
That is, we can still expand a two-parameter family of diffeomorphisms,
at order $n$, using Lie derivatives with respect to a finite number of
vector fields.

Then, here we are going to show how to approximate
a given family of diffeomorphisms $\Phi$, up to order $\lambda^k\Omega^{k'}$
with $k+k'=4$, in terms of some differential operators that we will show
later can be identified with Lie derivative operators with respect to
some vector field.  In other words, we are looking for an expansion of
the following type:
\beq
\fl \Phi^\ast_{\lambda,\Omega}T & & = \sum_{\begin{array}{c}k,k'=0 \\
k+k'=4\end{array}}^4 \lambda^k\Omega^{k'}\;(\mbox{Combination of
differential operators})^{}_{k,k'}\;T\,. \label{mainidea}
\eeq
We have performed this expansion term by term. For the sake of brevity
we present here only the most relevant developments and results.  We start by
introducing the set of differential operators that we have used to build
such an expansion by giving their actions on a general tensorial quantity
$T$:\footnote{The subscripts $(p,q)$ denote the lowest order in the
expansion~(\ref{mainidea}) at which these operators will appear for the
first time. See equation~(\ref{Taylorexpression}) below.}
\beq
\fl\cL_{(1,0)}T & := &
\left[\frac{\partial}{\partial\lambda}\Phi^*_{\lambda,\Omega}T\right]_{\lo}
\,, \label{operator10} \\
\fl\cL_{(0,1)}T & := &
\left[\frac{\partial}{\partial\Omega}\Phi^*_{\lambda,\Omega}T\right]_{\lo}
\,, \\
\fl\cL_{(2,0)}T & := &
\left[\frac{\partial^2}{\partial\lambda^2}\Phi^*_{\lambda,\Omega}T\right]_{
\lo}-\cL^2_{(1,0)}T\,, \\
\fl\cL_{(1,1)}T & := &
\left[\frac{\partial^2}{\partial\lambda\partial\Omega}
\Phi^*_{\lambda,\Omega}T\right]_{\lo}
\!\!\!\!\!-\left(\epsilon_0\cL_{(1,0)}\cL_{(0,1)}+
\epsilon_1\cL_{(0,1)}\cL_{(1,0)}
\right)T\label{L11}\,, \\
\fl\cL_{(0,2)}T & := &
\left[\frac{\partial^2}{\partial\Omega^2}\Phi^*_{\lambda,\Omega}T\right]_{
\lo}-\cL^2_{(0,1)}T\,, \\
\fl\cL_{(3,0)}T & := &
\left[\frac{\partial^3}{\partial\lambda^3}\Phi^*_{\lambda,\Omega}T\right]_{
\lo}-3\cL_{(1,0)}\cL_{(2,0)}T-\cL^3_{(1,0)}T\,, \\
\fl\cL_{(2,1)}T & := &
\left[\frac{\partial^3}{\partial\lambda^2\partial\Omega}
\Phi^*_{\lambda,\Omega}T\right]_{\lo}
\nn\\
\fl & &  -2\cL_{(1,0)}\cL_{{(1,1)}}T-\cL_{{(0,1)}}\cL_{{(2,0)}}T-
2\epsilon_2\cL_{{(1,0)}}\cL_{{(0,1)}}\cL_{{(1,0)}}T\nn\\
\fl & &  -(\epsilon_1-\epsilon_2)\cL_{{(0,1)}}\cL^2_{{(1,0)}}T
-(\epsilon_0-\epsilon_2)\cL^2_{{(1,0)}}\cL_{{(0,1)}}T
\label{L21}\,, \\
\fl\cL_{(1,2)}T & := &
\left[\frac{\partial^3}{\partial\lambda\partial\Omega^2}
\Phi^*_{\lambda,\Omega}T\right]_{\lo} \nn \\
\fl & &  -2\cL_{{(0,1)}}\cL_{{(1,1)}}T-\cL_{{(1,0)}}\cL_{{(0,2)}}T-
2\epsilon_3\cL_{{(0,1)}}\cL_{{(1,0)}}\cL_{{(0,1)}}T\nn\\
\fl & &  -(\epsilon_0-\epsilon_3)\cL_{{(1,0)}}\cL^2_{{(0,1)}}T
-(\epsilon_1-\epsilon_3)\cL^2_{{(0,1)}}\cL_{{(1,0)}}T
\label{L12}\,, \\
\fl \cL_{(0,3)}T & := &
\left[\frac{\partial^3}{\partial\Omega^3}\Phi^*_{\lambda,\Omega}T\right]_{
\lo}-3\cL_{(0,1)}\cL_{(0,2)}T-\cL^3_{(0,1)}T\,, \\
\fl \cL_{(4,0)}T & := &
\left[\frac{\partial^4}{\partial\lambda^4}\Phi^*_{\lambda,\Omega}T\right]_{
\lo}\nn\\
\fl & & -4\cL_{(1,0)}\cL_{(3,0)}T-3\cL^2_{(2,0)}T
-6\cL^2_{(1,0)}\cL_{(2,0)}T-\cL^4_{(1,0)}T\,, \\
\fl\cL_{(3,1)}T & := &
\left[\frac{\partial^4}{\partial\lambda^3\partial\Omega}
\Phi^*_{\lambda,\Omega}T\right]_{\lo} \nn \\
\fl & &  -3\cL_{{(1,0)}}\cL_{{(2,1)}}T
-\cL_{{(0,1)}}\cL_{{(3,0)}}T-3\epsilon_4\cL_{{(2,0)}}\cL_{{(1,1)}}T
-3\epsilon_5\cL_{{(1,1)}}\cL_{{(2,0)}}T\nn \\
\fl & &  -3\cL^2_{{(1,0)}}\cL_{{(1,1)}}T
-3\left(\epsilon_0\cL_{{(1,0)}}\cL_{{(0,1)}}T
+\epsilon_1\cL_{{(0,1)}}\cL_{{(1,0)}}\right)\cL_{{(2,0)}}T \nn \\
\fl & &  -(\epsilon_1-\epsilon_2-\epsilon_6)\cL_{{(0,1)}}\cL^3_{{(1,0)}}T
-3\epsilon_6\cL_{{(1,0)}}\cL_{{(0,1)}}\cL^2_{{(1,0)}}T\nn\\
\fl & &  -3(\epsilon_2-\epsilon_6)\cL^2_{{(1,0)}}\cL_{{(0,1)}}\cL_{{(1,0)}}T
-(\epsilon_0-2\epsilon_2+\epsilon_6)\cL^3_{{(1,0)}}\cL_{{(0,1)}}T \,, \\
\fl\cL_{(2,2)}T & := &
\left[\frac{\partial^4}{\partial\lambda^2\partial\Omega^2}
\Phi^*_{\lambda,\Omega}T\right]_{\lo}
\nn\\
\fl & &  -2\cL_{{(1,0)}}\cL_{{(1,2)}}T+2\cL_{{(0,1)}}\cL_{{(2,1)}}T-
2\cL^2_{{(1,1)}}T \nn \\
\fl & &  -\epsilon_7\cL_{{(2,0)}}\cL_{{(0,2)}}T
-\epsilon_8\cL_{{(0,2)}}\cL_{{(2,0)}}T -
\cL^2_{{(1,0)}}\cL_{{(0,2)}}T-\cL^2_{{(0,1)}}\cL_{{(2,0)}}T \nn \\
\fl & &  -4\left(\epsilon_0\cL_{{(1,0)}}\cL_{{(0,1)}}T+
\epsilon_1\cL_{{(0,1)}}\cL_{{(1,0)}}\right)\cL_{{(1,1)}}T \nn \\
\fl & &  +(\epsilon_3+\epsilon_2-\epsilon_1+\epsilon_9)
\cL^2_{{(0,1)}}\cL^2_{{(1,0)}}T
+(\epsilon_3+\epsilon_2-\epsilon_0-\epsilon_9)
\cL^2_{{(1,0)}}\cL^2_{{(0,1)}}T\nn\\
\fl & &  -2(\epsilon_3+\epsilon_2-\epsilon_0\epsilon_1-\epsilon_9)
\cL_{{(1,0)}}\cL_{{(0,1)}}\cL_{{(1,0)}}\cL_{{(0,1)}}T\nn\\
\fl & &  -2(\epsilon_3+\epsilon_2-\epsilon_0\epsilon_1+\epsilon_9)
\cL_{{(0,1)}}\cL_{{(1,0)}}\cL_{{(0,1)}}\cL_{{(1,0)}}T\nn\\
\fl & &  +2(\epsilon_3-\epsilon_0\epsilon_1)
\cL_{{(1,0)}}\cL^2_{{(0,1)}}\cL_{{(1,0)}}T
+2(\epsilon_2-\epsilon_0\epsilon_1)
\cL_{{(0,1)}}\cL^2_{{(1,0)}}\cL_{{(0,1)}}T \,, \\
\fl\cL_{(1,3)}T & := &
\left[\frac{\partial^4}{\partial\lambda\partial\Omega^3}
\Phi^*_{\lambda,\Omega}T\right]_{\lo}\nn\\
\fl & &  -3\cL_{{(0,1)}}\cL_{{(1,2)}}T
-\cL_{{(1,0)}}\cL_{{(0,3)}}T-3\epsilon_{10}\cL_{{(0,2)}}\cL_{{(1,1)}}T
-3\epsilon_{11}\cL_{{(1,1)}}\cL_{{(0,2)}}T\nn \\
\fl & &  -3\cL^2_{{(0,1)}}\cL_{{(1,1)}}T
-3\left(\epsilon_0\cL_{{(1,0)}}\cL_{{(0,1)}}T+
\epsilon_1\cL_{{(0,1)}}\cL_{{(1,0)}}\right)\cL_{{(0,2)}}T \nn \\
\fl & &  -(\epsilon_0-\epsilon_3-\epsilon_{12})\cL_{{(1,0)}}\cL^3_{{(0,1)}}T
-3\epsilon_{12}\cL_{{(0,1)}}\cL_{{(1,0)}}\cL^2_{{(0,1)}}T\nn\\
\fl & &  -3(\epsilon_3-\epsilon_{12})\cL^2_{{(0,1)}}\cL_{{(1,0)}}
\cL_{{(0,1)}}T-(\epsilon_1-2\epsilon_3+\epsilon_{12})\cL^3_{{(0,1)}}
\cL_{{(1,0)}}T \,, \\
\fl\cL_{(0,4)}T & := &
\left[\frac{\partial^4}{\partial\Omega^4}\Phi^*_{\lambda,\Omega}T\right]_{
\lo}\nn\\
\fl & &  -4\cL_{(0,1)}\cL_{(0,3)}T-3\cL^2_{(0,2)}T
-6\cL^2_{(0,1)}\cL_{(0,2)}T-\cL^4_{(0,1)}T\,, \label{operator04}
\eeq
where the quantities $\epsilon_A$ ($A=0,\dots,12$) are real numbers which
must satisfy the following conditions
\beq
\fl \epsilon_0+\epsilon_1 = 1\,,~~~
\epsilon_4+\epsilon_5 = 1\,, ~~~
\epsilon_7+\epsilon_8 = 1\,, ~~~
\epsilon_{10}+\epsilon_{11} = 1\,.  \label{constraints}
\eeq
These constants reflect the fact that, as already mentioned above,
there is not a unique way of constructing an expansion of the
type~(\ref{mainidea}), actually the number of possibilities is
infinite.  In this paper we have restricted ourselves
to expansions in which every single term has the form
$\cL_{(p_1,q_1)}\cdots\cL_{(p_n,q_n)}T$ with
$p_1+q_1 \leq \cdots \leq p_n+q_n$.  Then, the constants
$\epsilon_A$ express the freedom that we have in constructing
the expansion~(\ref{mainidea}) with this criteria.
This freedom and the non-uniqueness of the construction above is not a problem:
the operators $\cL_{(p,q)}$ and the corresponding vectors (see below) that we
are going to define should be though of as a ``basis" for the construction in
terms of Lie derivative of each of the terms in the expansion
(\ref{Taylorfamily}). This ``basis" is not unique, but the result at each order
is.

On the other hand, it is straightforward to check that these operators
are linear and satisfy the Leibnitz rule and hence they are derivatives.
We can also check that they satisfy the rest of conditions of the
theorem stated in~\ref{sebastiano}. Therefore, for each of them there is a
vector field $\xi_{(p,q)}$ such that
\be
\cL_{\xi_{(p,q)}}T := \cL_{(p,q)}T~~~~~ (p,q\in\NN)\,.
\label{lieder}
\ee

In the particular case when $\Phi$ is a group of diffeomorphisms  we
recover the previous case (subsection~\ref{groupdif}),
and $\cL_{(p,q)}=0$ if $p+q>1$.

Using the differential operators we have just introduced we can express
the Taylor expansion (\ref{Taylorfamily}) of $\Phi_{\lambda,\Omega}$,
up to fourth order in $\lambda$ and $\Omega$,
in terms of the Lie derivatives associated with the vector fields
$\xi_{(p,q)}$~(\ref{lieder}):
\beq
\fl \Phi^\ast_{\lambda,\Omega}T & & =  T + \lambda\cL_{\xi_{(1,0)}}T +
\Omega\cL_{\xi_{(0,1)}}T  \nn \\
\fl & & + \frac{\lambda^2}{2}\left\{\cL_{\xi_{(2,0)}}
+\cL^2_{\xi_{(1,0)}}\right\}T+\frac{\Omega^2}{2}\left\{\cL_{\xi_{(0,2)}}
+\cL^2_{\xi_{(0,1)}}\right\}T \nn \\
\fl & & + \lambda\Omega\left\{\cL_{\xi_{(1,1)}}+\epsilon_0
\cL_{\xi_{(1,0)}}\cL_{\xi_{(0,1)}}+\epsilon_1\cL_{\xi_{(0,1)}}
\cL_{\xi_{(1,0)}}\right\}T \nn \\
\fl & & + \frac{\lambda^3}{6}\left\{\cL_{\xi_{(3,0)}}+3\cL_{\xi_{(1,0)}}
\cL_{\xi_{(2,0)}}+\cL^3_{\xi_{(1,0)}}\right\}T \nn \\
\fl & & + \frac{\lambda^2\Omega}{2}\left\{\cL_{\xi_{(2,1)}}
+2\cL_{\xi_{(1,0)}}\cL_{\xi_{(1,1)}}+\cL_{\xi_{(0,1)}}\cL_{\xi_{(2,0)}}
+2\epsilon_2\cL_{\xi_{(1,0)}}\cL_{\xi_{(0,1)}}\cL_{\xi_{(1,0)}}\right. \nn \\
\fl & & \left. +(\epsilon_1-\epsilon_2)\cL_{\xi_{(0,1)}}\cL^2_{\xi_{(1,0)}}
+(\epsilon_0-\epsilon_2)\cL^2_{\xi_{(1,0)}}\cL_{\xi_{(0,1)}}\right\}T \nn \\
\fl & & + \frac{\lambda\Omega^2}{2}\left\{\cL_{\xi_{(1,2)}}+2
\cL_{\xi_{(0,1)}}\cL_{\xi_{(1,1)}}+\cL_{\xi_{(1,0)}}\cL_{\xi_{(0,2)}}
+2\epsilon_3\cL_{\xi_{(0,1)}}\cL_{\xi_{(1,0)}}
\cL_{\xi_{(0,1)}}\right. \nn \\
\fl & & \left. +(\epsilon_0-\epsilon_3)\cL_{\xi_{(1,0)}}\cL^2_{\xi_{(0,1)}}
+(\epsilon_1-\epsilon_3)\cL^2_{\xi_{(0,1)}}\cL_{\xi_{(1,0)}}\right\}T \nn \\
\fl & & + \frac{\Omega^3}{6}\left\{\cL_{\xi_{(0,3)}}+3\cL_{\xi_{(0,1)}}
\cL_{\xi_{(0,2)}}+\cL^3_{\xi_{(0,1)}}\right\}T \nn \\
\fl & & +
\frac{\lambda^4}{24}\left\{\cL_{\xi_{(4,0)}}+4\cL_{\xi_{(1,0)}}
\cL_{\xi_{(3,0)}}+3\cL^2_{\xi_{(2,0)}}+6\cL^2_{\xi_{(1,0)}}
\cL_{\xi_{(2,0)}}+\cL^4_{\xi_{(1,0)}}\right\}T  \nn \\
\fl & & +\frac{\lambda^3\Omega}{6}\left\{\cL_{\xi_{(3,1)}}+3\cL_{\xi_{(1,0)}}
\cL_{\xi_{(2,1)}}+\cL_{\xi_{(0,1)}}\cL_{\xi_{(3,0)}}+3\epsilon_4
\cL_{\xi_{(2,0)}}\cL_{\xi_{(1,1)}}\right. \nn \\
\fl & & +3\epsilon_5\cL_{\xi_{(1,1)}}\cL_{\xi_{(2,0)}}
+ 3\cL^2_{\xi_{(1,0)}}\cL_{\xi_{(1,1)}}
+3\left(\epsilon_0\cL_{\xi_{(1,0)}}\cL_{\xi_{(0,1)}}+
\epsilon_1\cL_{\xi_{(0,1)}}\cL_{\xi_{(1,0)}}\right)\cL_{\xi_{(2,0)}} \nn \\
\fl & & + (\epsilon_1-\epsilon_2-\epsilon_6)\cL_{\xi_{(0,1)}}
\cL^3_{\xi_{(1,0)}}+3\epsilon_6\cL_{\xi_{(1,0)}}\cL_{\xi_{(0,1)}}
\cL^2_{\xi_{(1,0)}} \nn \\
\fl & &
\left. +3(\epsilon_2-\epsilon_6)\cL^2_{\xi_{(1,0)}}\cL_{\xi_{(0,1)}}
\cL_{\xi_{(1,0)}}+(\epsilon_0-2\epsilon_2+\epsilon_6)\cL^3_{\xi_{(1,0)}}
\cL_{\xi_{(0,1)}}\right\}T \nn \\
\fl & & + \frac{\lambda^2\Omega^2}{4}\left\{
\cL_{\xi_{(2,2)}}+2\cL_{\xi_{(1,0)}}\cL_{\xi_{(1,2)}}+2\cL_{\xi_{(0,1)}}
\cL_{\xi_{(2,1)}}+2\cL^2_{\xi_{(1,1)}} \right.   \nn \\
\fl & & + \epsilon_7\cL_{\xi_{(2,0)}}\cL_{\xi_{(0,2)}}
+\epsilon_8\cL_{\xi_{(0,2)}}\cL_{\xi_{(2,0)}}+\cL^2_{\xi_{(1,0)}}
\cL_{\xi_{(0,2)}}+\cL^2_{\xi_{(0,1)}}\cL_{\xi_{(2,0)}} \nn \\
\fl & & + 4\left(\epsilon_0\cL_{\xi_{(1,0)}}\cL_{\xi_{(0,1)}}+
\epsilon_1\cL_{\xi_{(0,1)}}\cL_{\xi_{(1,0)}}\right)\cL_{\xi_{(1,1)}} \nn \\
\fl & & \left. -(\epsilon_3+\epsilon_2-\epsilon_1+\epsilon_9)
\cL^2_{\xi_{(0,1)}}\cL^2_{\xi_{(1,0)}}-(\epsilon_3+\epsilon_2-
\epsilon_0-\epsilon_9)\cL^2_{\xi_{(1,0)}}\cL^2_{\xi_{(0,1)}}\right. \nn \\
\fl & & +2(\epsilon_3+\epsilon_2-\epsilon_0\epsilon_1-\epsilon_9)
\cL_{\xi_{(1,0)}}\cL_{\xi_{(0,1)}}\cL_{\xi_{(1,0)}}\cL_{\xi_{(0,1)}} \nn \\
\fl & & +2(\epsilon_3+\epsilon_2-\epsilon_0\epsilon_1+\epsilon_9)
\cL_{\xi_{(0,1)}}\cL_{\xi_{(1,0)}}\cL_{\xi_{(0,1)}}\cL_{\xi_{(1,0)}} \nn \\
\fl & & \left.-2(\epsilon_3-\epsilon_0\epsilon_1)
\cL_{\xi_{(1,0)}}\cL^2_{\xi_{(0,1)}}\cL_{\xi_{(1,0)}}
-2(\epsilon_2-\epsilon_0\epsilon_1)
\cL_{\xi_{(0,1)}}\cL^2_{\xi_{(1,0)}}\cL_{\xi_{(0,1)}} \right\}T \nn \\
\fl & & + \frac{\lambda\Omega^3}{6}\left\{\cL_{\xi_{(1,3)}}
+3\cL_{\xi_{(0,1)}}\cL_{\xi_{(1,2)}}+\cL_{\xi_{(1,0)}}\cL_{\xi_{(0,3)}}
+3\epsilon_{10}\cL_{\xi_{(0,2)}}\cL_{\xi_{(1,1)}} \right. \nn \\
\fl & & +3\epsilon_{11}\cL_{\xi_{(1,1)}}\cL_{\xi_{(0,2)}}
+ 3\cL^2_{\xi_{(0,1)}}\cL_{\xi_{(1,1)}}
+3\left(\epsilon_0\cL_{\xi_{(1,0)}}\cL_{\xi_{(0,1)}}+
\epsilon_1\cL_{\xi_{(0,1)}}\cL_{\xi_{(1,0)}}\right)\cL_{\xi_{(0,2)}} \nn \\
\fl & & +(\epsilon_0-\epsilon_3-\epsilon_{12})\cL_{\xi_{(1,0)}}
\cL^3_{\xi_{(0,1)}}+3\epsilon_{12}\cL_{\xi_{(0,1)}}\cL_{\xi_{(1,0)}}
\cL^2_{\xi_{(0,1)}} \nn \\
\fl & & \left. +3(\epsilon_3-\epsilon_{12})\cL^2_{\xi_{(0,1)}}
\cL_{\xi_{(1,0)}}\cL_{\xi_{(0,1)}}+(\epsilon_1-2\epsilon_3+\epsilon_{12})
\cL^3_{\xi_{(0,1)}}\cL_{\xi_{(1,0)}}\right\}T \nn \\
\fl & & +
\frac{\Omega^4}{24}\left\{\cL_{\xi_{(0,4)}}+4\cL_{\xi_{(0,1)}}
\cL_{\xi_{(0,3)}}+3\cL^2_{\xi_{(0,2)}}+6\cL^2_{\xi_{(0,1)}}
\cL_{\xi_{(0,2)}}+\cL^4_{\xi_{(0,1)}}\right\}T \nn \\
\fl & & + O^5(\lambda,\Omega)\,. \label{Taylorexpression}
\eeq
In this expression we can see the way in which the parameters $\{\epsilon_A\}$
[real constants subject to the conditions~(\ref{constraints})]
describe the arbitrariness we have in the {\em reconstruction}
of the Taylor expansion of a two--parameter family of diffeomorphisms in
terms of Lie derivative operators.

To finish this section we will show how to  recover the
one-parameter case from the two-parameter case.
The case when one of $\lambda$ or $\Omega$ vanishes is trivial and it can be
recovered from the above expressions just setting either $\lambda=0$ or
$\Omega=0$. Let then consider the only other case, which arises when
the  two parameters $\lambda$ and $\Omega$
are not longer independent, e.g. $\Omega=\Omega(\lambda)$.
Then the specific way of recovering the single parameter ($\lambda$) case will
depend on the specific function $\Omega(\lambda)$. Here we illustrate the
simplest case  of a linear relation
$\Omega = a \lambda$ ($a\neq 0$).    To arrive to the one-parameter expansion
(see~\cite{BMMS,SB,BS}), we need to study the consequences of
the dependence between the two parameters.   We can do that by
looking at the definitions of the operators $\cL_{(p,q)}$ in equations
(\ref{operator10}-\ref{operator04}), and at their association with
Lie derivative operators, described by equation~(\ref{lieder}).
The result is a set of relations between the vector fields
$\xi_{(p,q)}$ that can be summarized in the following relation:
\be
\xi_{(p,q)} = \frac{1}{a^q}\xi_{(p+q,0)} \,.
\ee
Then, rescaling the vector fields $\xi_{(p,0)}$ in the following
way: $\xi_{(p,0)} \rightarrow (p+1)^{-1}\xi_{(p,0)}$, we arrive
at exactly the same expansions as in the one-parameter case~\cite{BMMS,SB,BS}.
  In the case in which
both parameters $\lambda$ and $\Omega$ have a specific  physical meaning,
it will be the physics that will impose the functional dependence
$\Omega(\lambda)$ in particular sub-cases.

%
%

\section{Gauges in perturbation theory and the two--parameter case}
\label{gauges}

\begin{figure}
\begin{center}
\includegraphics[height=4in,width=5in,bbllx=99, bblly=255,
bburx=595, bbury=650]{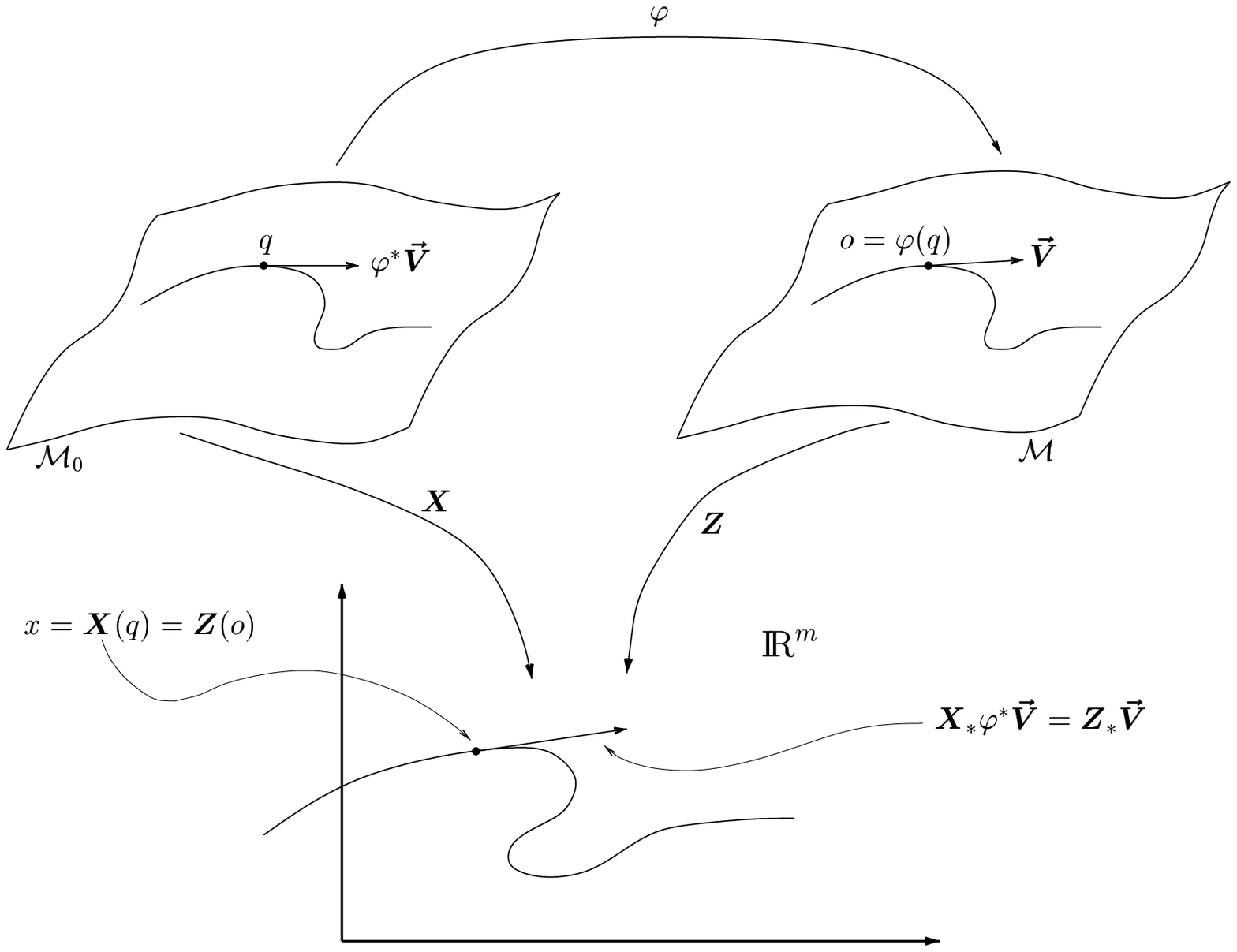}
\end{center}
\caption{By choosing the coordinates on $\cM_0$ and
$\cM$ in such a way that $(\mb{Z}\circ\varphi)^\mu
=x^\mu$, a curve in $\cM_0$, and its $\varphi$--transformed in
$\cM$ have the same representation in $\RR^m$.
Therefore, the components of the tangent vectors ${{V}}$ and
$\varphi^*V$ at the points $\varphi(p)$ and $p$ are the same:
$(\varphi^*V)^\mu(x)=(\varphi^*V)x^\mu\left.\right|_q=
V(\mb{X}\circ\varphi^{-1})^\mu\left.\right|_{\varphi(q)}=
V\,z^\mu\left.\right|_{\varphi(q)}= V^{\mu}(x)$.}
\label{fig1}
\end{figure}

Let us consider for the moment a spacetime $\{g^{\rm b},\cM_{0}\}$
which we call the background, and a physical spacetime  $\{g,\cM\}$ which we
attempt to describe as a perturbation of
$\{g^{\rm b},\cM_{0}\}$.\footnote{As manifolds $\cM_0$ and $\cM$ are the
same; for generality we assume that they are $m$--dimensional.}
In relativistic perturbation theory we are used to write expressions
of the form
\be
g_{\mu\nu}(x) = g^{\rm b}_{\mu\nu}(x) +\delta g_{\mu\nu}(x)\,,
\label{metricpert}
\ee
relating a perturbed tensor field such as the metric with the
background value of the same field and with the perturbation. In
doing this, we are implicitly assigning a correspondence between
points of the perturbed and the background spacetimes~\cite{bi:schutz}.
Indeed through (\ref{metricpert}), which is a relation between the
images of the fields in ${\rm I\! R}^{m}$ rather than between the fields
themselves on the respective manifolds $\cM$ and $\cM_{0}$, we are
saying that there is a unique point $x$ in ${\rm I\! R}^{m}$ that
is at the same time the image of  {\it two} points: one (say $q$) in
$\cM_{0}$ and one ($o$) in $\cM$.
This correspondence is what is usually called a gauge choice in the
context of perturbation theory. Clearly, this is more than the usual
assignment of coordinate labels to points of a single
spacetime~\cite{sachs}.  Furthermore, the correspondence established by
relations such as (\ref{metricpert}) is not \mbox{\em per se} unique,
but rather (\ref{metricpert}) typically defines a set of gauges,
unless certain specific restrictions are satisfied by the fields
involved (e.g.\,, some metric components vanish). Leaving this problem
aside, i.e.\ supposing that the gauge has been somehow completely fixed,
let us look more precisely at the implications of (\ref{metricpert}),
adopting the geometrical description illustrated in Figure~\ref{fig1}.

If we call $\mb{X}$ the chart on $\cM_0$ and $\mb{Z}$ the chart on
$\cM$ we see that if we choose $z(o)=x(q)$, i.e.\ the correspondence between
points of $\{g^{\rm b},\cM_{0}\}$ and $\{g,\cM\}$ implicit
in~(\ref{metricpert}), we are implicitly defining a map $\varphi$
between $\cM_0$ and $\cM$, such that $\varphi=\mb{Z}^{-1}\circ\mb{X}$.
Thus from the geometrical point of view a gauge choice is an
identification of points of $\cM_0$ and $\cM$.  Therefore, we could as
well start directly assigning the point  identification map $\varphi$
first, calling $\varphi$ itself a gauge, and
defining coordinates adapted to it later.  This turns out to be
a simpler way of proceeding in order to derive the gauge
transformations in the following section.

Let us follow this idea in the specific case of two parameters
introducing, in the spirit of~\cite{stewart,BMMS,SB,BS}, an $(m+2)$--dimensional
manifold $\cN$, foliated by $m$--dimensional submanifolds diffeomorphic to
$\cM$, so that $\cN=\cM\times\RR^2$.  We shall label each copy of
$\cM$ by the corresponding value of the parameters
$\lambda,\Omega$.  The manifold $\cN$ has a natural differentiable
structure which is the direct product of those of $\cM$ and
$\RR^2$.  We can then choose charts on $\cN$ in which $x^{\mu}$
($\mu=0,1,\dots m-1$) are coordinates of each leaf
$\cM_{\lambda,\Omega}$ and $x^m=\lambda,\,x^{m+1}=\Omega$.

In this construction
we are assuming that the perturbed spacetimes have the same manifold
as the background one.  With this we are not allowing the possibility
of addressing questions like how can perturbations affect the differential
structure of the background spacetime.  These issues would require
the use of a much more complicated mathematical apparatus, in particular
the notion of limits of spacetimes introduced by Geroch~\cite{GEROCH}
in order to define the background manifold as a limit
$\lambda,\Omega\rightarrow 0$ of a family of {\em perturbed} manifolds
$\cM_{\lambda,\Omega}$.  Instead, we consider perturbations as fields
{\em living} on the background (as in~\cite{stewart,BMMS,SB,BS}), a standard 
approach in which these issues do not appear.  More sofisticated structures 
of the extended manifold $\cN$ have
been considered in~\cite{MAIA}, in an attempt to give the background
metric the status of a gauge invariant quantity.

Coming back to our construction, if a tensor $T_{\lambda,\Omega}$ is given
on each $\cM_{\lambda,\Omega}$, we have that a tensor field $T$ is
automatically defined on $\cN$ by the relation
$T(p,\lambda,\Omega):= T_{\lambda,\Omega}(p)$, with
$p\in\cM_{\lambda,\Omega}$.\footnote{Tensor
fields on $\cN$ constructed in this way are
``tangent'' to $\cM$, in the sense that their components
$m$ and $m+1$ in the charts we have defined vanish
identically.}
In particular, on each $\cM_{\lambda,\Omega}$ one has a
metric $g_{\lambda,\Omega}$ and a set of matter fields
$\tau_{\lambda,\Omega}$, satisfying the set of field equations
\be
{\cal E}[g_{\lambda,\Omega},\tau_{\lambda,\Omega}]=0\,.
\ee
Correspondingly, the fields $g$ and $\tau$ are defined on
$\cN$.

We now want to define the perturbation in any tensor $T$,
therefore we must find a way to compare
$T_{\lambda,\Omega}$ with $T_0$: this requires a prescription for
identifying points of $\cM_{\lambda,\Omega}$ with those of
$\cM_0$. This is easily accomplished by assigning a
diffeomorphism
$\varphi_{\lambda,\Omega}:\cN\rightarrow\cN$ such that
$\left.\varphi_{\lambda,\Omega}\right|_{\cM_0}:\cM_0
\rightarrow\cM_{\lambda,\Omega}$.
Clearly, $\varphi_{\lambda,\Omega}$ can be regarded as the
member of a two--parameter group of diffeomorphisms $\varphi$ on $\cN$, 
corresponding to the values of $\lambda,\Omega$ of the group
parameter. Therefore, we could equally well give the
vector fields $^{\varphi}\eta,\,^{\varphi}\zeta$ that generate
$\varphi$. In the chart introduced above,
$^{\varphi}\eta^m=1$, $^{\varphi}\eta^{m+1}=0$,
$^{\varphi}\zeta^m=0$,  $^{\varphi}\zeta^{m+1}=1$ but, except for
these conditions, $^{\varphi}\eta,\,^{\varphi}\zeta$ remain
arbitrary.  For convenience, we shall also refer to such a pair 
of vector fields as a gauge. It is always possible to take the 
chart above defined such that $\,^{\varphi}\eta^{\mu}=\,^{\varphi}
\zeta^{\mu}=0$.  So, in this chart, point of different 
submanifolds $\cM_{\lambda,\Omega}$ connected by the
diffeomorphism $\varphi$ have the same $\cM$--coordinates
$x^0,\dots,x^{m-1}$, and
differ only by the value of the coordinates $\lambda,\,\Omega$.
We call such a chart ``adapted to the gauge $\varphi$'': this is 
what is always used in practice.

The perturbation in $T$ can now be defined simply as
\be
\Delta^{\varphi}_0
T^\varphi_{\lambda,\Omega} := \left.\varphi^*_{\lambda,\Omega}T\right|_{\cM_0}
-T_0\,.  \label{defperturbation}
\ee
The first term on the right--hand side of
(\ref{defperturbation}) can be Taylor--expanded using (\ref{Taylorgroup})
to get
\be
\Delta^{\varphi}_0 T^\varphi_{\lambda,\Omega}=\sum_{k,k'=0}^{\infty}
\frac{\lambda^k{\Omega^{k}}'}{k!\,k'!}\delta^{(k,k')}_\varphi T-T_0\,,
\label{perturbationexpansion}
\ee
where
\be
\delta^{(k,k')}_\varphi T:=\left[\frac{\partial^{k+k'}}
{\partial \lambda^k\partial\Omega^{k'}}\varphi^*_{\lambda,\Omega}T
\right]_{\lambda=0,\Omega=0,\cM_0}\!\!\! = \left.\cL^k_{\,^{\varphi}\!\eta}
\cL^{k'}_{\,^{\varphi}\!\zeta}T\right|_{\cM_0}\,,
\label{defperturbationk}
\ee
which defines the perturbation of order $(k,k')$ of 
$T$ (notice that $\delta^{(0,0)}_\varphi T=T_0$).  It is worth noticing
$\Delta_0 T^\varphi_{\lambda,\Omega}$ and $\delta^{(k,k')}_\varphi T$ are
defined on $\cM_0$; this formalizes the statement one
commonly finds in the literature that ``perturbations are
fields living in the background''. It is important to
appreciate that the parameters $\lambda,\,\Omega$
labelling the various spacetime models also serve to
perform the expansion (\ref{perturbationexpansion}), and
therefore determine what one means by ``perturbations of
order $(k,k')$''.

%
%

\section{Gauge invariance and gauge transformations}
\label{gaugeiandt}
Let us now suppose that two gauges $\varphi$ and $\psi$,
described by pairs of vector fields $(^{\varphi}\eta,\,^{\varphi}\zeta)$
and $(^{\psi}\eta,\,^{\psi}\zeta)$ respectively, are
defined on $\cN$, such that in the chart discussed 
above\footnote{In general, if the chart is adapted to the gauge $\varphi$, 
i.e.\,
$\,^{\varphi}\eta^{\mu}=\,^{\varphi}\zeta^{\mu}=0$, it is not adapted
to the gauge $\psi$, so 
$\,^{\psi}\eta^{\mu}\neq 0,\,^{\psi}\zeta^{\mu}\neq0$.}
\be
\begin{array}{lclcl}
^{\varphi}\eta^{m} & = & \,^{\psi}\eta^m & = & 1 \,,  \\
^{\varphi}\zeta^{m}& = & \,^{\psi}\zeta^{m} & = & 0 \,, \\
^{\varphi}\eta^{m+1} & = & \,^{\psi}\eta^{m+1} & = & 0 \,, \\
^{\varphi}\zeta^{m+1} & = & \,^{\psi}\zeta^{m+1} & = & 1 \,. \label{1010}
\end{array}
\ee
Correspondingly, the integral curves of 
$(^{\varphi}\eta,\,^{\varphi}\zeta)$ and $(^{\psi}\eta,\,^{\psi}\zeta)$
define two two--parameter groups of diffeomorphisms $\varphi$ and
$\psi$ on $\cN$, that connect any two leaves of the
foliation. Thus, $(^{\varphi}\eta,\,^{\varphi}\zeta)$
and $(^{\psi}\eta,\,^{\psi}\zeta)$ are everywhere transverse to
$\cM_{\lambda,\Omega}$ and points lying on the same
integral surface of either of the two are to be regarded
{\it as the same point} within the respective gauge:
$\varphi$ and $\psi$ are both point identification maps,
i.e.\ two different gauge choices.

The pairs of vector fields $(^{\varphi}\eta,\,^{\varphi}\zeta)$
and $(^{\psi}\eta,\,^{\psi}\zeta)$ can both be used to pull back a
generic tensor $T$ and therefore to construct two other
tensor fields $\varphi^*_{\lambda,\Omega}T$ and
$\psi^*_{\lambda,\Omega}T$, for any given value of
$(\lambda,\Omega)$. In particular, on $\cM_0$ we now have
three tensor fields, i.e.\ $T_0$ and
\be
T^\varphi_{\lambda,\Omega}:= \left.\varphi^*_{\lambda,\Omega}
T\right|_{\cM_0}\,,~~~~~~~~
T^\psi_{\lambda,\Omega}:= \left.\psi^*_{\lambda,\Omega}
T\right|_{\cM_0}\,.
\ee
Since $\varphi$ and $\psi$ represent gauge choices for mapping a perturbed 
manifold $\cM_{\lambda,\Omega}$ into the unperturbed one $\cM_0$,
$T^\varphi_{\lambda,\Omega}$ and
$T^\psi_{\lambda,\Omega}$ are the representations, in $\cM_0$, of the perturbed
tensor
according to the two gauges. We can write, 
using~(\ref{defperturbation})--(\ref{defperturbationk}) and the 
expansion~(\ref{Taylorgroup}),
\beq
\fl T^\varphi_{\lambda,\Omega} & = &
\sum_{k=0}^{\infty}\frac{\lambda^k{\Omega^k}'}{k!\,k'!}
\delta^{(k,k')}_\varphi T=\sum_{k,k'=0}^{\infty}
\frac{\lambda^k{\Omega^k}'}{k!\,k'!}
\cL^k_{\,^{\varphi\!}\eta}\cL^{k'}_{\,^{\varphi\!}\zeta}T=T_0+
\Delta_0^{\varphi}
T_{\lambda,\Omega}\,,\label{defexpX}\\
\fl T^\psi_{\lambda,\Omega} & = &
\sum_{k=0}^{\infty}\frac{\lambda^k{\Omega^k}'}{k!\,k'!}
\delta^{(k,k')}_\psi T = \sum_{k,k'=0}^{\infty}
\frac{\lambda^k{\Omega^k}'}{k!\,k'!}
\cL^k_{\,^{\psi\!}\eta}\cL^{k'}_{\,^{\psi\!}\zeta}T=T_0+\Delta_0^{\psi}
T_{\lambda,\Omega}\label{defexpY}
\eeq
where $\delta^{(k,k')}_\varphi T$, $\delta^{(k,k')}_\psi T$ are the perturbations
(\ref{defperturbationk}) in the gauges $\varphi$ and $\psi$ respectively, i.e.
\beq
\delta^{(k,k')}_\varphi T & = & \left. \cL^k_{\,^{\varphi}\!\eta}
\cL^{k'}_{\,^{\varphi}\!\zeta}T\right|_{\cM_0} \,, \\
\delta^{(k,k')}_\psi T & = & \left. \cL^k_{\,^{\psi}\!\eta} 
\cL^{k'}_{\,^{\psi}\!\zeta}T\right|_{\cM_0}\,.
\eeq

\subsection{Gauge invariance}\label{gaugeinvariance}

If $T^\varphi_{\lambda,\Omega}=T^\psi_{\lambda,\Omega}$, for any pair of gauges 
$\varphi$ and $\psi$, we
say that $T$ is {\em totally gauge invariant}. This is a very strong
condition, because 
then (\ref{defexpX}) and (\ref{defexpY}) imply that $\delta^{(k,k')}_\varphi T=
\delta^{(k,k')}_\psi T$, for all gauges $\varphi$ and $\psi$ and for any 
$(k,k')$.  In any
practical case, however, one is interested in perturbations up to a fixed
order. It is thus convenient to weaken the definition above, saying
that $T$ is {\em gauge invariant up to order $(n,n')$}
iff for any two gauges $\varphi$ and $\psi$
\be
\delta^{(k,k')}_\varphi T =\delta^{(k,k')}_\psi T
~~~~~~\forall~(k,k')~~\hbox{ with}~ k\leq n\,,\,k'\leq n'\,.\label{charatGI}
\ee

We have that  a tensor field $T$ is gauge invariant to order $(n,n')$ iff
in a given gauge $\varphi$ we have that $\cL_{\xi}\delta^{(k,k')}_\varphi T=0$,
for any vector field $\xi$ defined on $\cM$ and for any $(k,k')<(n,n')$.

To prove this statement, let us first show that it
is true for $(n,n')=(1,0)$. In fact, if
$\delta^{(1,0)}_\varphi T = \delta^{(1,0)}_\psi T$ for
two arbitrary gauges $\varphi,\psi$, we have
$\cL_{\,^{\varphi}\!\eta-^{\psi}\!\eta}T|_{\cM_0}
=0$. But since $\varphi$ and $\psi$ are arbitrary gauges, it follows that
$\,^{\varphi}\!\eta-\,^{\psi}\!\eta$ is an arbitrary
field $\xi$, and $\xi^m=\xi^{m+1}=0$ because
$\,^{\varphi}\!\eta^m=\,{}^{\psi}\!\eta^m=1,\,
\,^{\varphi}\!\eta^{m+1}=\,^{\psi}\!\eta^{m+1}=0$, so
$\xi$ is tangent to $\cM$. In the same way one proves the statement for
$(n,n')=(0,1)$. 
Now let us suppose that the statement is true for some $(n,n')$. Then, if
one also has 
$\delta^{(n+1,n')}_\varphi T|_{\cM_0}=\delta^{(n+1,n')}_\psi T|_{\cM_0}$, it follows
that 
$\cL_{\,^{\varphi}\!\eta-^{\psi}\!\eta}\delta^{(n,n')}_\varphi T=0$,
while if $\delta^{(n,n'+1)}_\varphi T|_{\cM_0}=\delta^{(n,n'+1)}_\psi T|_{\cM_0}$,
it follows that 
$\cL_{\,^{\varphi}\!\zeta-^{\psi}\!\zeta}\delta^{(n,n')}_\varphi T=0$,
and we establish the result by induction over $(n,n')$.

As a consequence, $T$ is gauge invariant to order $(n,n')$ iff $T_0$ and all
its
perturbations of order lower than $(n,n')$ are, in any gauge, either
vanishing or 
constant scalars, or a combination of Kronecker deltas with constant
coefficients. Thus,
this generalizes to an arbitrary order $(n,n')$ and to the two--parameter
case the results of \cite{BMMS,sachs,stewart}. Further, it then follows that
$T$ is 
totally gauge invariant iff it is a combination of Kronecker deltas with
coefficients
depending only on $\lambda,\Omega$.

\subsection{Gauge transformations}

If a tensor $T$ is not gauge invariant, it is important to know how its
representation
on $\cM_0$ changes under a gauge transformation. To this purpose it is 
natural to introduce, 
for each value of $(\lambda,\Omega)\in\RR^2$, the diffeomorphism
$\Phi_{\lambda,\Omega}
:\cM_0\rightarrow\cM_0$ defined by
\be
\Phi_{\lambda,\Omega}:=\varphi_{-\!\lambda,-\!\Omega}\circ\psi_{\lambda,
\Omega}\,.
\ee
Given that from the geometrical point of view adopted here $\varphi$ and 
$\psi$ are two gauges, $\Phi$ represents the gauge transformation.
\begin{figure}
\begin{center}
\includegraphics[height=4in,width=5in,bbllx=99, bblly=255,
bburx=595, bbury=650]{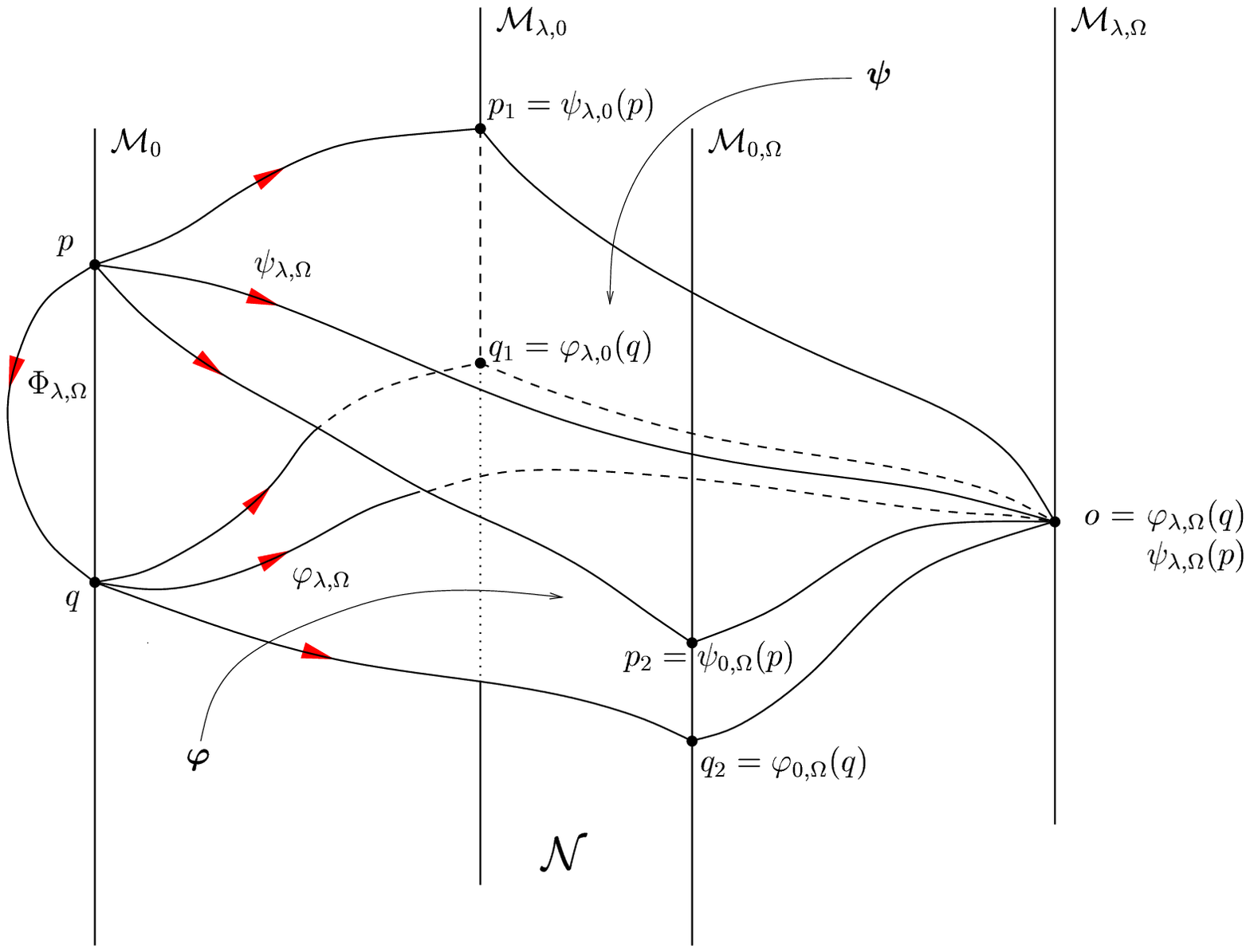}
\end{center}
\caption{The action of a gauge transformation $\Phi_{\lambda,\Omega}$,
represented on the background spacetime $\cM_0$. The gauge $\psi$
is a two-dimensional submanifold embedded in $\cN=\cM\times\RR^2$,
represented by the surface connecting the points $p$, $p_1$, $p_2$, and
$o$.  Similarly, the gauge $\varphi$ is represented by the surface
connecting the points $q$, $q_1$, $q_2$, and $o$.}
\label{fig:GT}
\end{figure}
The action of $\Phi_{\lambda,\Omega}$ is illustrated in Fig.\ \ref{fig:GT}.
We must stress that $\Phi:\cM_0\times\RR^2\rightarrow\cM_0$ thus defined,
{\em is not} 
a two--parameter group of diffeomorphisms in $\cM_0$. In fact,
$\Phi_{\lambda_1,\Omega_1}\circ\Phi_{\lambda_2,\Omega_2}\neq
\Phi_{\lambda_1+\lambda_2,
\Omega_1+\Omega_2}$, essentially because the fields 
$(^{\varphi}\eta,\,^{\varphi}\zeta)$ and $(^{\psi}\eta,\,^{\psi}\zeta)$ have, 
in general, a 
non--vanishing commutator. However, it can be Taylor expanded, using the
results of
section (\ref{familydiffeo}).

The tensor fields $T^\varphi_{\lambda,\Omega}$ and $T^\psi_{\lambda,\Omega}$, 
defined on $\cM_0$ by the gauges $\varphi$ and $\psi$, are connected 
by the linear map $\Phi_{\lambda,\Omega}^*$:
\beq
T^\psi_{\lambda,\Omega} & = & \left. 
\psi^*_{\lambda,\Omega}T\,\right|_{\cM_0}= \left.
\left(\psi^*_{\lambda,\Omega}\varphi^*_{\!-\!\lambda,\!-\!\Omega}
\varphi^*_{\lambda,\Omega}T\right)\right|_{\cM_0} \nn \\
& = & \left. \Phi^*_{\lambda,\Omega}\left(\varphi^*_{\lambda,\Omega}T
\right)\right|_{\cM_0} = \Phi^*_{\lambda,\Omega}T^\varphi_{\lambda,\Omega}\,.
\eeq
Thus, the gauge transformation to an arbitrary order $(n,n')$ is given
by the Taylor expansion of the pull--back 
$\Phi^*_{\lambda,\Omega}T$, whose terms are explicitly given in
section~\ref{familydiffeo}.  Up to fourth order, 
we have explicitly from (\ref{Taylorexpression})
\beq
\fl T^\psi_{\lambda,\Omega} & & = T^\varphi_{\lambda,\Omega}
+\lambda\cL_{\xi_{(1,0)}}T^\varphi_{\lambda,\Omega}+
\Omega\cL_{\xi_{(0,1)}}T^\varphi_{\lambda,\Omega} \nn \\
\fl & & + \frac{\lambda^2}{2}\left\{\cL_{\xi_{(2,0)}}
+\cL^2_{\xi_{(1,0)}}\right\}T^\varphi_{\lambda,\Omega}
+\frac{\Omega^2}{2}\left\{\cL_{\xi_{(0,2)}}+\cL^2_{\xi_{(0,1)}}\right\}
T^\varphi_{\lambda,\Omega} \nn \\
\fl & & + \lambda\Omega\left\{\cL_{\xi_{(1,1)}}
+\epsilon_0\cL_{\xi_{(1,0)}}\cL_{\xi_{(0,1)}}+
\epsilon_1\cL_{\xi_{(0,1)}}\cL_{\xi_{(1,0)}}\right\}
T^\varphi_{\lambda,\Omega} \nn \\
\fl & & + \frac{\lambda^3}{6}\left\{\cL_{\xi_{(3,0)}}+3\cL_{\xi_{(1,0)}}
\cL_{\xi_{(2,0)}}+\cL^3_{\xi_{(1,0)}}\right\}
T^\varphi_{\lambda,\Omega} \nn \\ 
\fl & & + \frac{\lambda^2\Omega}{2}\left\{\cL_{\xi_{(2,1)}}
+2\cL_{\xi_{(1,0)}}\cL_{\xi_{(1,1)}}+\cL_{\xi_{(0,1)}}\cL_{\xi_{(2,0)}}
+2\epsilon_2\cL_{\xi_{(1,0)}}\cL_{\xi_{(0,1)}}
\cL_{\xi_{(1,0)}}\right. \nn \\
\fl & & \left.+(\epsilon_1-\epsilon_2)\cL_{\xi_{(0,1)}}\cL^2_{\xi_{(1,0)}}
+(\epsilon_0-\epsilon_2)\cL^2_{\xi_{(1,0)}}\cL_{\xi_{(0,1)}}\right\}
T^\varphi_{\lambda,\Omega} \nn \\ 
\fl & & + \frac{\lambda\Omega^2}{2}
\left\{\cL_{\xi_{(1,2)}}+2\cL_{\xi_{(0,1)}}\cL_{\xi_{(1,1)}}
+\cL_{\xi_{(1,0)}}\cL_{\xi_{(0,2)}}+2\epsilon_3\cL_{\xi_{(0,1)}}
\cL_{\xi_{(1,0)}}\cL_{\xi_{(0,1)}}\right. \nn \\
\fl & & \left.+(\epsilon_0-\epsilon_3)\cL_{\xi_{(1,0)}}\cL^2_{\xi_{(0,1)}}
+(\epsilon_1-\epsilon_3)\cL^2_{\xi_{(0,1)}}\cL_{\xi_{(1,0)}}\right\}
T^\varphi_{\lambda,\Omega} \nn \\ 
\fl & & + \frac{\Omega^3}{6}\left\{\cL_{\xi_{(0,3)}}+3\cL_{\xi_{(0,1)}}
\cL_{\xi_{(0,2)}}+\cL^3_{\xi_{(0,1)}}\right\}
T^\varphi_{\lambda,\Omega} \nn \\
\fl & & + \frac{\lambda^4}{24}\left\{\cL_{\xi_{(4,0)}}
+4\cL_{\xi_{(1,0)}}\cL_{\xi_{(3,0)}}
+3\cL^2_{\xi_{(2,0)}}+6\cL^2_{\xi_{(1,0)}}\cL_{\xi_{(2,0)}}
+\cL^4_{\xi_{(1,0)}}\right\}T^\varphi_{\lambda,\Omega} \nn \\ 
\fl & & + \frac{\lambda^3\Omega}{6}\left\{\cL_{\xi_{(3,1)}} 
+3\cL_{\xi_{(1,0)}}\cL_{\xi_{(2,1)}}
+\cL_{\xi_{(0,1)}}\cL_{\xi_{(3,0)}}+3\epsilon_4\cL_{\xi_{(2,0)}}
\cL_{\xi_{(1,1)}} \right. \nn \\
\fl & & +3\epsilon_5\cL_{\xi_{(1,1)}}
\cL_{\xi_{(2,0)}} + 3\cL^2_{\xi_{(1,0)}}\cL_{\xi_{(1,1)}} 
+3\left(\epsilon_0\cL_{\xi_{(1,0)}}\cL_{\xi_{(0,1)}}+
\epsilon_1\cL_{\xi_{(0,1)}}\cL_{\xi_{(1,0)}}\right)\cL_{\xi_{(2,0)}} \nn \\
\fl & & +(\epsilon_1-\epsilon_2-\epsilon_6)\cL_{\xi_{(0,1)}}
\cL^3_{\xi_{(1,0)}}+3\epsilon_6\cL_{\xi_{(1,0)}}\cL_{\xi_{(0,1)}}
\cL^2_{\xi_{(1,0)}} \nn \\
\fl & & \left. +3(\epsilon_2-\epsilon_6)\cL^2_{\xi_{(1,0)}}
\cL_{\xi_{(0,1)}}\cL_{\xi_{(1,0)}}+(\epsilon_0-2\epsilon_2+\epsilon_6)
\cL^3_{\xi_{(1,0)}}\cL_{\xi_{(0,1)}}\right\}T^\varphi_{\lambda,\Omega} \nn \\ 
\fl & & + \frac{\lambda^2\Omega^2}{4}
\left\{\cL_{\xi_{(2,2)}}+2\cL_{\xi_{(1,0)}}\cL_{\xi_{(1,2)}}
+2\cL_{\xi_{(0,1)}}\cL_{\xi_{(2,1)}}+
2\cL^2_{\xi_{(1,1)}} \right.   \nn \\
\fl & & + \epsilon_7\cL_{\xi_{(2,0)}}\cL_{\xi_{(0,2)}} 
+\epsilon_8\cL_{\xi_{(0,2)}}\cL_{\xi_{(2,0)}} + 
\cL^2_{\xi_{(1,0)}}\cL_{\xi_{(0,2)}}
+\cL^2_{\xi_{(0,1)}}\cL_{\xi_{(2,0)}} \nn \\ 
\fl & & + 4\left(\epsilon_0\cL_{\xi_{(1,0)}}\cL_{\xi_{(0,1)}}+
\epsilon_1\cL_{\xi_{(0,1)}}\cL_{\xi_{(1,0)}}\right)\cL_{\xi_{(1,1)}} \nn \\
\fl & & \left. -(\epsilon_3+\epsilon_2-\epsilon_1+\epsilon_9)
\cL^2_{\xi_{(0,1)}}\cL^2_{\xi_{(1,0)}}
-(\epsilon_3+\epsilon_2-\epsilon_0-\epsilon_9)
\cL^2_{\xi_{(1,0)}}\cL^2_{\xi_{(0,1)}}\right. \nn \\
\fl & & +2(\epsilon_3+\epsilon_2-\epsilon_0\epsilon_1-\epsilon_9)
\cL_{\xi_{(1,0)}}\cL_{\xi_{(0,1)}}\cL_{\xi_{(1,0)}}\cL_{\xi_{(0,1)}} \nn \\
\fl & & +2(\epsilon_3+\epsilon_2-\epsilon_0\epsilon_1+\epsilon_9)
\cL_{\xi_{(0,1)}}\cL_{\xi_{(1,0)}}\cL_{\xi_{(0,1)}}\cL_{\xi_{(1,0)}} \nn \\
\fl & & \left.-2(\epsilon_3-\epsilon_0\epsilon_1)
\cL_{\xi_{(1,0)}}\cL^2_{\xi_{(0,1)}}\cL_{\xi_{(1,0)}}
-2(\epsilon_2-\epsilon_0\epsilon_1)\cL_{\xi_{(0,1)}}\cL^2_{\xi_{(1,0)}}
\cL_{\xi_{(0,1)}} \right\}T^\varphi_{\lambda,\Omega} \nn \\
\fl & & + \frac{\lambda\Omega^3}{6}\left\{\cL_{\xi_{(1,3)}}
+3\cL_{\xi_{(0,1)}}\cL_{\xi_{(1,2)}}+\cL_{\xi_{(1,0)}}\cL_{\xi_{(0,3)}}
+3\epsilon_{10}\cL_{\xi_{(0,2)}}\cL_{\xi_{(1,1)}} 
\right. \nn \\
\fl & & +3\epsilon_{11}\cL_{\xi_{(1,1)}}\cL_{\xi_{(0,2)}} 
+3\cL^2_{\xi_{(0,1)}}\cL_{\xi_{(1,1)}} 
+3\left(\epsilon_0\cL_{\xi_{(1,0)}}\cL_{\xi_{(0,1)}}+
\epsilon_1\cL_{\xi_{(0,1)}}\cL_{\xi_{(1,0)}}\right)\cL_{\xi_{(0,2)}} \nn \\
\fl & & +(\epsilon_0-\epsilon_3-\epsilon_{12})\cL_{\xi_{(1,0)}}
\cL^3_{\xi_{(0,1)}}+3\epsilon_{12}\cL_{\xi_{(0,1)}}\cL_{\xi_{(1,0)}}
\cL^2_{\xi_{(0,1)}} \nn \\
\fl & & \left. +3(\epsilon_3-\epsilon_{12})\cL^2_{\xi_{(0,1)}}
\cL_{\xi_{(1,0)}}\cL_{\xi_{(0,1)}}+(\epsilon_1-2\epsilon_3
+\epsilon_{12})\cL^3_{\xi_{(0,1)}}\cL_{\xi_{(1,0)}}
\right\}T^\varphi_{\lambda,\Omega} \nn \\ 
\fl & & + \frac{\Omega^4}{24}\left\{\cL_{\xi_{(0,4)}}
+4\cL_{\xi_{(0,1)}}\cL_{\xi_{(0,3)}}+3\cL^2_{\xi_{(0,2)}}
+6\cL^2_{\xi_{(0,1)}}\cL_{\xi_{(0,2)}}+\cL^4_{\xi_{(0,1)}}\right\}
T^\varphi_{\lambda,\Omega} \nn \\ 
\fl & & +  O^5(\lambda,\Omega) \,,  \label{totalgaugetransf}
\eeq
where the $\xi_{(p,q)}$ are now the generators of 
the gauge transformation $\Phi_{\lambda,\Omega}$.

We can now relate the perturbations in the two gauges. To order $(n,n')$
with $n+n'\leq 4$, these relations can be derived by 
substituting (\ref{defexpX}), (\ref{defexpY}) in 
(\ref{totalgaugetransf}):
\beq
\fl \delta^{(1,0)}_\psi T-\delta^{(1,0)}_\varphi T = \cL_{\xi_{(1,0)}}T_0\,, \label{gi10}
\eeq
\beq
\fl \delta^{(0,1)}_\psi T-\delta^{(0,1)}_\varphi T = \cL_{\xi_{(0,1)}}T_0\,, \label{gi01}
\eeq
\beq
\fl \delta^{(2,0)}_\psi T-\delta^{(2,0)}_\varphi T = 2\cL_{\xi_{(1,0)}} 
\delta^{(1,0)}_\varphi T
+\left\{\cL_{\xi_{(2,0)}}+\cL_{\xi_{(1,0)}}^2\right\}T_0\,, \label{gi20}
\eeq
\beq
\fl \delta^{(1,1)}_\psi T-\delta^{(1,1)}_\varphi T = \cL_{\xi_{(1,0)}} 
\delta^{(0,1)}_\varphi T +
\cL_{\xi_{(0,1)}}\delta_\varphi^{(1,0)}T  \nn \\
\fl ~~~~~ + \left\{\cL_{\xi_{(1,1)}}+\epsilon_0\cL_{\xi_{(1,0)}}
\cL_{\xi_{(0,1)}}+\epsilon_1\cL_{\xi_{(0,1)}}\cL_{\xi_{(1,0)}}
\right\}T_0 \,,  \label{gi11}
\eeq
\beq
\fl \delta^{(0,2)}_\psi T-\delta^{(0,2)}_\varphi T = 2\cL_{\xi_{(0,1)}} 
\delta^{(0,1)}_\varphi T
+\left\{\cL_{\xi_{(0,2)}}+\cL_{\xi_{(0,1)}}^2\right\}T_0\,, \label{gi02}
\eeq
\beq
\fl \delta^{(3,0)}_\psi T-\delta^{(3,0)}_\varphi T = 3\cL_{\xi_{(1,0)}}
\delta_\varphi^{(2,0)}T
+ 3\left\{\cL_{\xi_{(2,0)}}+\cL^2_{\xi_{(1,0)}}\right\}\delta_\varphi^{(1,0)}T
\nn \\
\fl ~~~~~ + \left\{\cL_{\xi_{(3,0)}}+3\cL_{\xi_{(1,0)}}\cL_{\xi_{(2,0)}}
+\cL^3_{\xi_{(1,0)}}\right\}T_0 \,, \label{gi30}
\eeq
\beq
\fl \delta^{(2,1)}_\psi T-\delta^{(2,1)}_\varphi T = 2\cL_{\xi_{(1,0)}} 
\delta_\varphi^{(1,1)}T
+ \cL_{\xi_{(0,1)}}\delta_\varphi^{(2,0)}T + \left\{\cL_{\xi_{(2,0)}}
+ \cL^2_{\xi_{(1,0)}}\right\}\delta_\varphi^{(0,1)}T \nn \\
\fl ~~~~~ + 2\left\{\cL_{\xi_{(1,1)}}+\epsilon_0\cL_{\xi_{(1,0)}}
\cL_{\xi_{(0,1)}}+\epsilon_1\cL_{\xi_{(0,1)}}\cL_{\xi_{(1,0)}}\right\}
\delta_\varphi^{(1,0)}T \nn \\
\fl ~~~~~ + \left\{\cL_{\xi_{(2,1)}}+2\cL_{\xi_{(1,0)}}\cL_{\xi_{(1,1)}}
+\cL_{\xi_{(0,1)}}\cL_{\xi_{(2,0)}}+2\epsilon_2\cL_{\xi_{(1,0)}}
\cL_{\xi_{(0,1)}} \cL_{\xi_{(1,0)}}\right. \nn \\
\fl ~~~~~ \left.+(\epsilon_1-\epsilon_2)\cL_{\xi_{(0,1)}}\cL^2_{\xi_{(1,0)}}
+(\epsilon_0-\epsilon_2)\cL^2_{\xi_{(1,0)}}\cL_{\xi_{(0,1)}}\right\}T_0\,, 
\label{gi21}
\eeq
\beq
\fl \delta^{(1,2)}_\psi T-\delta^{(1,2)}_\varphi T = 2\cL_{\xi_{(0,1)}} 
\delta_\varphi^{(1,1)}T
+ \cL_{\xi_{(1,0)}}\delta_\varphi^{(0,2)}T+
\left\{\cL_{\xi_{(0,2)}}+\cL^2_{\xi_{(0,1)}}\right\}\delta_\varphi^{(1,0)}T \nn \\
\fl ~~~~~ + 2\left\{\cL_{\xi_{(1,1)}}+\epsilon_0\cL_{\xi_{(1,0)}}
\cL_{\xi_{(0,1)}}+\epsilon_1\cL_{\xi_{(0,1)}}\cL_{\xi_{(1,0)}}\right\}
\delta_\varphi^{(0,1)}T \nn \\
\fl ~~~~~ + \left\{\cL_{\xi_{(1,2)}}+2\cL_{\xi_{(0,1)}}\cL_{\xi_{(1,1)}}
+\cL_{\xi_{(1,0)}}\cL_{\xi_{(0,2)}}+2\epsilon_3\cL_{\xi_{(0,1)}}
\cL_{\xi_{(1,0)}}\cL_{\xi_{(0,1)}}\right. \nn \\
\fl ~~~~~ \left.+(\epsilon_0-\epsilon_3)\cL_{\xi_{(1,0)}}\cL^2_{\xi_{(0,1)}}
+(\epsilon_1-\epsilon_3)\cL^2_{\xi_{(0,1)}}\cL_{\xi_{(1,0)}}\right\}T_0 \,, 
\label{gi12}
\eeq 
\beq
\fl \delta^{(0,3)}_\psi T-\delta^{(0,3)}_\varphi T = 3\cL_{\xi_{(0,1)}} 
\delta_\varphi^{(0,2)}T
+ 3\left\{\cL_{\xi_{(0,2)}}+\cL^2_{\xi_{(0,1)}}\right\}\delta_\varphi^{(0,1)}T \nn \\
\fl ~~~~~ + \left\{\cL_{\xi_{(0,3)}}+3\cL_{\xi_{(0,1)}}
\cL_{\xi_{(0,2)}}+\cL^3_{\xi_{(0,1)}}\right\}T_0\,, \label{gi03}
\eeq
\beq
\fl \delta^{(4,0)}_\psi T-\delta^{(4,0)}_\varphi T =  4\cL_{\xi_{(1,0)}}
\delta_\varphi^{(3,0)}T + 6\left\{\cL_{\xi_{(2,0)}}+\cL^2_{\xi_{(1,0)}}\right\}
\delta_\varphi^{(2,0)}T \nn \\
\fl ~~~~~ + 4\left\{\cL_{\xi_{(3,0)}}+3\cL_{\xi_{(1,0)}}\cL_{\xi_{(2,0)}}
+\cL^3_{\xi_{(1,0)}}\right\}\delta_\varphi^{(1,0)}T \nn \\ 
\fl ~~~~~ + \left\{\cL_{\xi_{(4,0)}}+4\cL_{\xi_{(1,0)}}\cL_{\xi_{(3,0)}}
+3\cL^2_{\xi_{(2,0)}}+6\cL^2_{\xi_{(1,0)}}\cL_{\xi_{(2,0)}}
+\cL^4_{\xi_{(1,0)}}\right\}T_0 \,, \label{gi40}
\eeq
\beq
\fl \delta^{(3,1)}_\psi T-\delta^{(3,1)}_\varphi T = 3\cL_{\xi_{(1,0)}} 
\delta_\varphi^{(2,1)}T
+\cL_{\xi_{(0,1)}}\delta_\varphi^{(3,0)}T + 3\left\{\cL_{\xi_{(2,0)}}
+\cL^2_{\xi_{(1,0)}}\right\}\delta_\varphi^{(1,1)}T \nn \\
\fl ~~~~~ + 3\left\{\cL_{\xi_{(1,1)}}+\epsilon_0\cL_{\xi_{(1,0)}}
\cL_{\xi_{(0,1)}}+\epsilon_1\cL_{\xi_{(0,1)}}\cL_{\xi_{(1,0)}}\right\}
\delta_\varphi^{(2,0)}T \nn \\
\fl ~~~~~ + \left\{\cL_{\xi_{(3,0)}}+3\cL_{\xi_{(1,0)}}
\cL_{\xi_{(2,0)}}+\cL^3_{\xi_{(1,0)}}\right\}\delta_\varphi^{(0,1)}T \nn \\ 
\fl ~~~~~ + 3\left\{\cL_{\xi_{(2,1)}}+2\cL_{\xi_{(1,0)}}\cL_{\xi_{(1,1)}}
+\cL_{\xi_{(0,1)}}\cL_{\xi_{(2,0)}}+2\epsilon_2\cL_{\xi_{(1,0)}}
\cL_{\xi_{(0,1)}}\cL_{\xi_{(1,0)}}\right. \nn \\
\fl ~~~~~ \left.+(\epsilon_1-\epsilon_2)\cL_{\xi_{(0,1)}}\cL^2_{\xi_{(1,0)}}
+(\epsilon_0-\epsilon_2)\cL^2_{\xi_{(1,0)}}\cL_{\xi_{(0,1)}}\right\}
\delta_\varphi^{(1,0)}T \nn \\ 
\fl ~~~~~ + \left\{\cL_{\xi_{(3,1)}}+ 3\cL_{\xi_{(1,0)}}
\cL_{\xi_{(2,1)}}+\cL_{\xi_{(0,1)}}\cL_{\xi_{(3,0)}}+3\epsilon_4
\cL_{\xi_{(2,0)}}\cL_{\xi_{(1,1)}} +3\epsilon_5\cL_{\xi_{(1,1)}}
\cL_{\xi_{(2,0)}}\right. \nn \\
\fl ~~~~~ + 3\cL^2_{\xi_{(1,0)}}\cL_{\xi_{(1,1)}} 
+3\left(\epsilon_0\cL_{\xi_{(1,0)}}\cL_{\xi_{(0,1)}}+
\epsilon_1\cL_{\xi_{(0,1)}}\cL_{\xi_{(1,0)}}\right)\cL_{\xi_{(2,0)}} \nn \\
\fl ~~~~~ + (\epsilon_1-\epsilon_2-\epsilon_6)\cL_{\xi_{(0,1)}}
\cL^3_{\xi_{(1,0)}} + 3\epsilon_6\cL_{\xi_{(1,0)}}\cL_{\xi_{(0,1)}}
\cL^2_{\xi_{(1,0)}} \nn \\
\fl ~~~~~ \left. + 3(\epsilon_2-\epsilon_6)\cL^2_{\xi_{(1,0)}}
\cL_{\xi_{(0,1)}}\cL_{\xi_{(1,0)}}+(\epsilon_0-2\epsilon_2+\epsilon_6)
\cL^3_{\xi_{(1,0)}}\cL_{\xi_{(0,1)}}\right\}T_0 \,, \label{gi31}
\eeq
\beq
\fl \delta^{(2,2)}_\psi T-\delta^{(2,2)}_\varphi T = 2\cL_{\xi_{(1,0)}} 
\delta_\varphi^{(1,2)}T
+2\cL_{\xi_{(0,1)}}\delta_\varphi^{(2,1)}T  \nn \\
\fl ~~~~~ + \left\{\cL_{\xi_{(2,0)}}+\cL^2_{\xi_{(1,0)}}\right\}
\delta_\varphi^{(0,2)}T +
\left\{\cL_{\xi_{(0,2)}}+\cL^2_{\xi_{(0,1)}}\right\}
\delta_\varphi^{(2,0)}T \nn \\
\fl ~~~~~ + 4\left\{\cL_{\xi_{(1,1)}}+\epsilon_0\cL_{\xi_{(1,0)}}
\cL_{\xi_{(0,1)}}+\epsilon_1\cL_{\xi_{(0,1)}}\cL_{\xi_{(1,0)}}\right\}
\delta_\varphi^{(1,1)}T \nn \\
\fl ~~~~~ + 2\left\{\cL_{\xi_{(2,1)}}+2\cL_{\xi_{(1,0)}}\cL_{\xi_{(1,1)}}
+\cL_{\xi_{(0,1)}}\cL_{\xi_{(2,0)}}+2\epsilon_2\cL_{\xi_{(1,0)}}
\cL_{\xi_{(0,1)}}\cL_{\xi_{(1,0)}}\right. \nn \\
\fl ~~~~~ \left.+(\epsilon_1-\epsilon_2)\cL_{\xi_{(0,1)}}\cL^2_{\xi_{(1,0)}}
+(\epsilon_0-\epsilon_2)\cL^2_{\xi_{(1,0)}}\cL_{\xi_{(0,1)}}\right\}
\delta_\varphi^{(0,1)}T \nn \\ 
\fl ~~~~~ + 2\left\{\cL_{\xi_{(1,2)}}+2\cL_{\xi_{(0,1)}}\cL_{\xi_{(1,1)}}
+\cL_{\xi_{(1,0)}}\cL_{\xi_{(0,2)}}+2\epsilon_3\cL_{\xi_{(0,1)}}
\cL_{\xi_{(1,0)}}\cL_{\xi_{(0,1)}}\right. \nn \\
\fl ~~~~~ \left.+(\epsilon_0-\epsilon_3)\cL_{\xi_{(1,0)}}\cL^2_{\xi_{(0,1)}}
+(\epsilon_1-\epsilon_3)\cL^2_{\xi_{(0,1)}}\cL_{\xi_{(1,0)}}\right\}
\delta_\varphi^{(1,0)}T \nn \\ 
\fl ~~~~~ +
\left\{\cL_{\xi_{(2,2)}}+2\cL_{\xi_{(1,0)}}\cL_{\xi_{(1,2)}}
+2\cL_{\xi_{(0,1)}}\cL_{\xi_{(2,1)}}+2\cL^2_{\xi_{(1,1)}} \right. \nn \\
\fl ~~~~~ +\epsilon_7\cL_{\xi_{(2,0)}}\cL_{\xi_{(0,2)}} 
+\epsilon_8\cL_{\xi_{(0,2)}}\cL_{\xi_{(2,0)}} + \cL^2_{\xi_{(1,0)}}
\cL_{\xi_{(0,2)}}+\cL^2_{\xi_{(0,1)}}\cL_{\xi_{(2,0)}} \nn \\
\fl ~~~~~ +4\left(\epsilon_0\cL_{\xi_{(1,0)}}\cL_{\xi_{(0,1)}}+
\epsilon_1\cL_{\xi_{(0,1)}}\cL_{\xi_{(1,0)}}\right)\cL_{\xi_{(1,1)}} \nn \\
\fl ~~~~~ \left. -(\epsilon_3+\epsilon_2-\epsilon_1+\epsilon_9)
\cL^2_{\xi_{(0,1)}}\cL^2_{\xi_{(1,0)}}
-(\epsilon_3+\epsilon_2-\epsilon_0-\epsilon_9)
\cL^2_{\xi_{(1,0)}}\cL^2_{\xi_{(0,1)}}\right. \nn \\
\fl ~~~~~ +2(\epsilon_3+\epsilon_2-\epsilon_0\epsilon_1-\epsilon_9)
\cL_{\xi_{(1,0)}}\cL_{\xi_{(0,1)}}\cL_{\xi_{(1,0)}}\cL_{\xi_{(0,1)}} \nn \\
\fl ~~~~~ +2(\epsilon_3+\epsilon_2-\epsilon_0\epsilon_1+\epsilon_9)
\cL_{\xi_{(0,1)}}\cL_{\xi_{(1,0)}}\cL_{\xi_{(0,1)}}\cL_{\xi_{(1,0)}} \nn \\
\fl ~~~~~ \left.-2(\epsilon_3-\epsilon_0\epsilon_1)
\cL_{\xi_{(1,0)}}\cL^2_{\xi_{(0,1)}}\cL_{\xi_{(1,0)}}
-2(\epsilon_2-\epsilon_0\epsilon_1)\cL_{\xi_{(0,1)}}\cL^2_{\xi_{(1,0)}}
\cL_{\xi_{(0,1)}} \right\}T_0\,, \label{gi22}
\eeq
\beq
\fl \delta^{(1,3)}_\psi T-\delta^{(1,3)}_\varphi T = 3\cL_{\xi_{(0,1)}} 
\delta_\varphi^{(1,2)}T
+ \cL_{\xi_{(1,0)}}\delta_\varphi^{(0,3)}T+ 
3\left\{\cL_{\xi_{(0,2)}}+\cL^2_{\xi_{(0,1)}}\right\}\delta_\varphi^{(1,1)}T \nn \\
\fl ~~~~~ + 3\left\{\cL_{\xi_{(1,1)}}+\epsilon_0\cL_{\xi_{(1,0)}}
\cL_{\xi_{(0,1)}}+\epsilon_1\cL_{\xi_{(0,1)}}\cL_{\xi_{(1,0)}}\right\}
\delta_\varphi^{(0,2)}T \nn \\
\fl ~~~~~ + \left\{\cL_{\xi_{(0,3)}}+3\cL_{\xi_{(0,1)}}
\cL_{\xi_{(0,2)}}+\cL^3_{\xi_{(0,1)}}\right\}\delta_\varphi^{(1,0)}T \nn \\
\fl ~~~~~ + 3\left\{\cL_{\xi_{(1,2)}}+2\cL_{\xi_{(0,1)}}\cL_{\xi_{(1,1)}}
+\cL_{\xi_{(1,0)}}\cL_{\xi_{(0,2)}}+2\epsilon_3\cL_{\xi_{(0,1)}}
\cL_{\xi_{(1,0)}}\cL_{\xi_{(0,1)}}\right. \nn \\
\fl ~~~~~ \left.+(\epsilon_0-\epsilon_3)\cL_{\xi_{(1,0)}}\cL^2_{\xi_{(0,1)}}
+(\epsilon_1-\epsilon_3)\cL^2_{\xi_{(0,1)}}\cL_{\xi_{(1,0)}}\right\}
\delta_\varphi^{(0,1)}T \nn \\ 
\fl ~~~~~ + \left\{\cL_{\xi_{(1,3)}}+3\cL_{\xi_{(0,1)}}\cL_{\xi_{(1,2)}}
+\cL_{\xi_{(1,0)}}\cL_{\xi_{(0,3)}}+3\epsilon_{10}\cL_{\xi_{(0,2)}}
\cL_{\xi_{(1,1)}}+3\epsilon_{11}\cL_{\xi_{(1,1)}}
\cL_{\xi_{(0,2)}}\right. \nn \\
\fl ~~~~~ + 3\cL^2_{\xi_{(0,1)}}\cL_{\xi_{(1,1)}} 
+3\left(\epsilon_0\cL_{\xi_{(1,0)}}\cL_{\xi_{(0,1)}}+
\epsilon_1\cL_{\xi_{(0,1)}}\cL_{\xi_{(1,0)}}\right)\cL_{\xi_{(0,2)}} \nn \\
\fl ~~~~~ +(\epsilon_0-\epsilon_3-\epsilon_{12})\cL_{\xi_{(1,0)}}
\cL^3_{\xi_{(0,1)}}+3\epsilon_{12}\cL_{\xi_{(0,1)}}\cL_{\xi_{(1,0)}}
\cL^2_{\xi_{(0,1)}} \nn \\
\fl ~~~~~ \left.+3(\epsilon_3-\epsilon_{12})\cL^2_{\xi_{(0,1)}}
\cL_{\xi_{(1,0)}}\cL_{\xi_{(0,1)}}+(\epsilon_1-2\epsilon_3+\epsilon_{12})
\cL^3_{\xi_{(0,1)}}\cL_{\xi_{(1,0)}}\right\}T_0\,, \label{gi13}
\eeq
\beq
\fl \delta^{(0,4)}_\psi T-\delta^{(0,4)}_\varphi T = 4\cL_{\xi_{(0,1)}}
\delta_\varphi^{(0,3)}T+6\left\{\cL_{\xi_{(0,2)}}+\cL^2_{\xi_{(0,1)}}\right\}
\delta_\varphi^{(0,2)}T \nn \\
\fl ~~~~~ + 4\left\{\cL_{\xi_{(0,3)}}+3\cL_{\xi_{(0,1)}}\cL_{\xi_{(0,2)}}
+\cL^3_{\xi_{(0,1)}}\right\}\delta_\varphi^{(0,1)}T \nn \\ 
\fl ~~~~~ + \left\{\cL_{\xi_{(0,4)}}+4\cL_{\xi_{(0,1)}}\cL_{\xi_{(0,3)}}
+3\cL^2_{\xi_{(0,2)}}+6\cL^2_{\xi_{(0,1)}}\cL_{\xi_{(0,2)}}
+\cL^4_{\xi_{(0,1)}}\right\}T_0\,. \label{gi04}
\eeq
This result is, of course, consistent with the characterization
of gauge invariance given in subsection~\ref{gaugeinvariance}. 
Equations (\ref{gi10}) and (\ref{gi01}) imply that
$T_{\lambda,\Omega}$
is gauge invariant to the order $(1,0)$ or $(0,1)$ iff
$\cL_{\xi}T_0=0$,
for any vector field on $\cM_0$. Equation (\ref{gi20}) implies that
$T_{\lambda,\Omega}$
is gauge invariant to the order $(2,0)$ iff $\cL_{\xi}T_0=0$ {\em and}
$\cL_{\xi}\delta^{(1,0)}_\varphi T=0$, for any vector field on $\cM_0$, and so 
on for all the orders.

It is also possible to find the explicit expressions for the generators 
$\xi_{(p,q)}$ of the gauge transformation $\Phi$ 
in terms of the gauge vector fields $({}^\varphi\eta,{}^\varphi\zeta)$
and $({}^\psi\eta,{}^\psi\zeta)$.   We write here their 
expressions up to second order:
\beq
\xi_{(1,0)} &=& {}^\psi\eta - {}^\varphi\eta \,, \\
\xi_{(0,1)} &=& {}^\psi\zeta - {}^\varphi\zeta \,, \\
\xi_{(2,0)} &=& [{}^\varphi\eta,{}^\psi\eta] \,, \\
\xi_{(1,1)} &=& \epsilon_0[{}^\varphi\eta,{}^\psi\zeta]+
\epsilon_1[{}^\varphi\zeta,{}^\psi\eta] \,, \\
\xi_{(0,2)} &=& [{}^\varphi\zeta,{}^\psi\zeta]\,.
\eeq
  
\subsection{Coordinate transformations}  
  
Up to now, we have built a two--parameter formalism using a
geometrical, coordinate--free language.  However, in order to carry
out explicit calculations in a practical case, one has to introduce 
systems of local coordinates.  In this respect, all our expressions
are immediately translated into components simply by using the 
expression of the components of the Lie derivative of a tensor.  
Nonetheless, much of the literature on the subject is written using 
coordinate systems, and gauge transformations are most often
represented by the corresponding coordinate transformations.
For this reason, we devote this subsection to describe how to
establish the translation between the two languages, giving in 
particular the explicit transformation of coordinates (further 
details are in \cite{MMB}).

\begin{figure}
\begin{center}
\includegraphics[height=3.4in,width=5in,bbllx=99, bblly=255,
bburx=600, bbury=560]{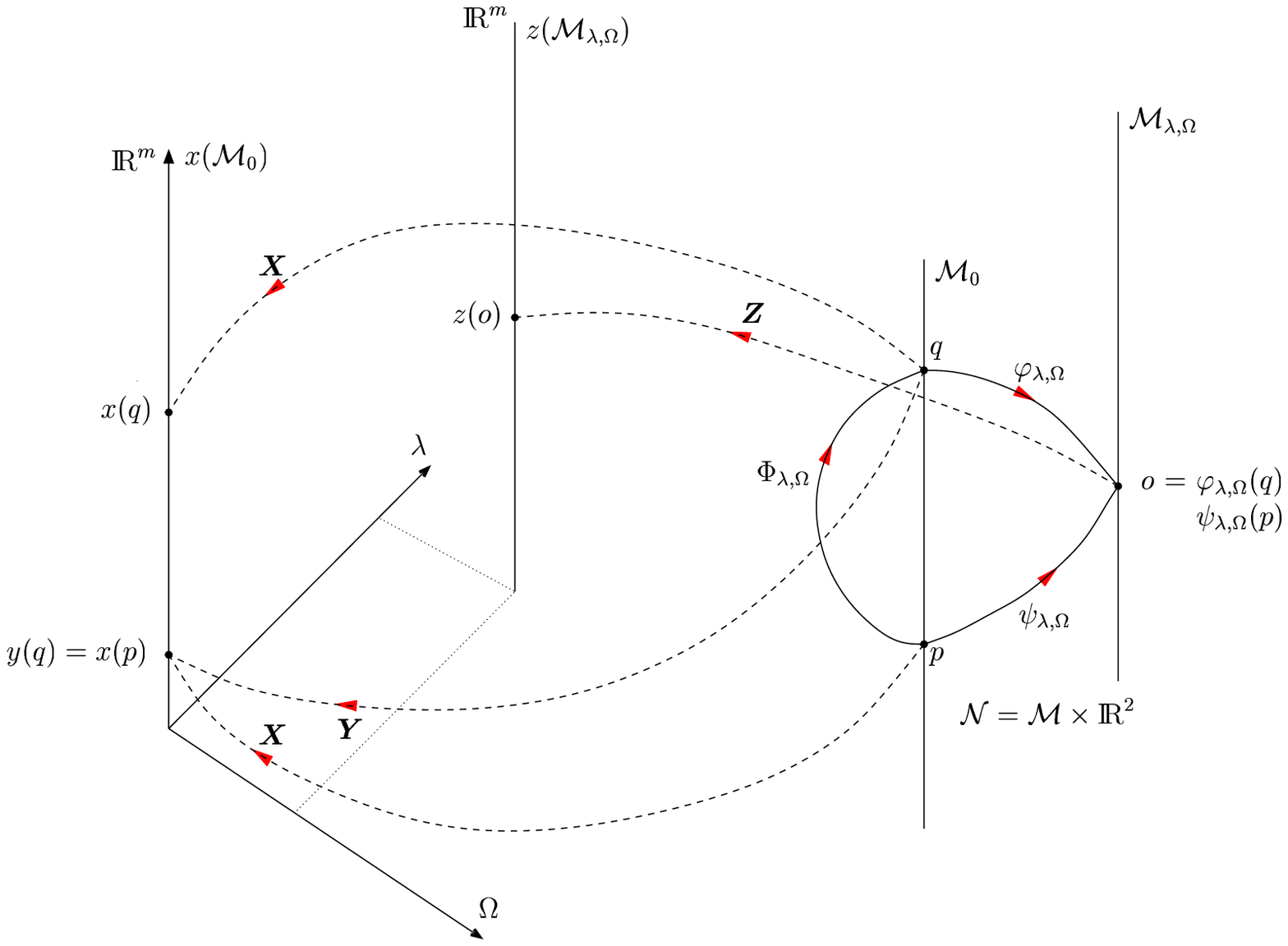}
\end{center}
\caption{The picture shows how two different gauge choices, $\varphi$ and $\psi$,
map two different points in the background manifold $\cM_0$, namely $q$ and
$p$ respectively, to the same point $o$ in the physical spacetime
$\cM_{\lambda\Omega}$.
The coordinate version of this fact is shown on
the left part of picture.  We can look at this either from the {\it active}
point of view (the two points $q$ and $p$ are given coordinates 
in the same chart by the map $\mb{X}$, namely $x^{\mu}(q)$ and 
$x^{\mu}(p)$), or from the {\it passive} point of 
view (the point $q$ is given coordinates in two different charts
by maps $\mb{X}$ and $\mb{Y}$, corresponding to two different points
of $\RR^m$, $x(q)$ and $y(q)=x(p)$). Finally, $\mb{Z}$ assigns
coordinates to the point $o$ of the physical spacetime $\cM_{\lambda\Omega}$.}
\label{fig:GTcoord}
\end{figure}

Let us consider the situation described in Fig.~\ref{fig:GTcoord}.
We have considered two gauge choices, represented by the groups of
diffeomorphisms $\varphi$ and $\psi$, under which the point $o$ in the
physical manifold $\cM_{\lambda,\Omega}$
corresponds to two different points in the background manifold
$\cM_0$, namely $q=\varphi^{-1}_{\lambda,\Omega}(o)$ and
$p=\psi_{\lambda,\Omega}^{-1}(o)$.  The transformation relating
these two gauge choices is described by the two--parameter
family of diffeomorphisms $\Phi_{\lambda,\Omega}=
\varphi^{-1}_{\lambda,\Omega}\circ\psi_{\lambda,\Omega}$, so that
$\Phi_{\lambda,\Omega}(p)=q$.  This gauge transformation maps
a tensor field $T$ on $q\in\cM_0$ to the tensor field
$(\Phi^*T)(p)=\Phi^*(T(q))$ on $p\in\cM_0$.

Now, let us consider a chart $({\cal U},\mb{X})$ on an open subset
${\cal U}$ of $\cM_0$.  The gauges $\varphi_{\lambda,\Omega}$ and 
$\psi_{\lambda,\Omega}$ define two maps from 
$\cM_{\lambda,\Omega}$ to $\RR^m$:
\beq  
\fl \begin{array}{lccc}
\mb{X}\circ\varphi^{-1}_{\lambda,\Omega}\,:&  
\cM_{\lambda,\Omega}&\longrightarrow &\RR^m\\  
& o &|\!\!\!\longrightarrow&x(q(o))\,, \\  
\end{array} 
\hspace{4mm}
\begin{array}{lccc}  
\mb{X}\circ\psi^{-1}_{\lambda,\Omega}\,:&  
\cM_{\lambda,\Omega}&\longrightarrow &\RR^m\\  
& o &|\!\!\!\longrightarrow&x(p(o))\,.\\  
\end{array}   
\eeq  
Then, we can look at the gauge transformation $\Phi$ in two
different ways: from the {\it active} point of view or from the
{\it passive} point of view.  In the first case, one considers
a diffeomorphism which changes the point on the background $\cM_0$. 
To these points one associates different values of the coordinates
in the chart $({\cal U},\mb{X})$.  So the coordinate change is
given by 
\be
x^{\mu}(p)~\longrightarrow~x^{\mu}(q) 
\ee  
or, defining the pull--back of $x$ as $\tilde{x}^{\mu}(p):=x^{\mu}(\Phi(p))$, 
\be
x^{\mu}(p)~\longrightarrow~{\tilde x}^{\mu}(p)\,.
\ee  
If we instead consider the passive point of view, we need 
to introduce a new chart $({\cal U}',y)$: 
\be
\mb{Y}:= \mb{X}\circ\Phi_{\lambda,\Omega}^{-1}\,, 
\ee  
such that the two sets of coordinates are related by
\be
y^{\mu}(q)=x^{\mu}(p) \,, 
\ee  
so we can say that the gauge transformation does not  change the   
point on $\cM_0$, but it changes the chart from $({\cal U},\mb{X})$ to   
$({\cal U}',\mb{Y})$, i.e.\ the labels of the point of $\cM_0$.  
The coordinate transformation is then  
\be
x^{\mu}(q)~\longrightarrow~y^{\mu}(q)\,.  
\ee
Now, let us consider the transformation of a vector field
$V$.  From the active point of view, the components
of $V$ in the chart $({\cal U},\mb{X})$, $V^\mu$,
are related with the ones of the transformed vector field 
${\tilde V}$, ${\tilde V}^{\mu}$, by
\be
{\tilde V}^{\mu}= (\mb{X}_*\tilde{V})^\mu = 
 (\mb{X}_*\Phi_{\lambda,\Omega}^*V)^\mu \,.
\ee
From the passive point of view, we can use the properties relating  
the pull--back and push--forward maps associated with diffeomorphisms:  
\be
\mb{X}_*\Phi_{\lambda,\Omega}^*V=
\mb{X}_*\Phi^{-1}_{*\lambda,\Omega}V=
\mb{Y}_*V\,,  
\ee  
so we get the well known result that the components of the transformed vector 
${\tilde V}$ in the coordinate system $\mb{X}$ are defined in terms of the
components of the vector $V$ in the new coordinate system $\mb{Y}$: 
\be  
{\tilde V}^{\mu}(x(p))= (\mb{Y}_*V(q))^{\mu}=
V^{\prime\mu}(y(q))=
\left. \left(\frac{\partial y^{\mu}}
{\partial x^{\nu}}\right)\right|_{x(q)}V^{\nu}(x(q))\,.
\label{vectorGTcoords}
\ee

In order to write down explicit expressions, we will apply the
expansion of the pull--back of $\Phi^*$ [See
equation~(\ref{Taylorexpression})] to the coordinate functions
$x^{\mu}$.  Then, the {\it active} coordinate transformation is
given by
\beq
\fl {\tilde x}^{\mu}(p)&=&x^{\mu}(q) = (\Phi^*x^{\mu})(p) \nn \\
\fl & = & x^{\mu}(p)+\lambda\xi^{\mu}_{(1,0)}+\Omega\xi^{\mu}_{(0,1)}
\nn\\
\fl & + & \frac{\lambda^2}{2}\left(\xi^{\mu}_{(2,0)}+\xi^{\nu}_{(1,0)}
\xi^{\mu}_{(1,0),\nu}\right)+\frac{\Omega^2}{2}\left(\xi^{\mu}_{(0,2)}
+\xi^{\nu}_{(0,1)}\xi^{\mu}_{(0,1),\nu}\right)  \nn \\
\fl & + & \lambda\Omega\left(\xi^{\mu}_{(1,1)}+\epsilon_0\xi^{\nu}_{(1,0)}
\xi^{\mu}_{(0,1),\nu}+\epsilon_1\xi^{\nu}_{(0,1)}\xi^{\mu}_{(1,0),\nu}\right) 
\nn \\
\fl & + & \frac{\lambda^3}{6}\left(\xi^{\mu}_{(3,0)}+3\xi^{\nu}_{(1,0)}
\xi^{\mu}_{(2,0),\nu}+\xi^{\rho}_{(1,0)}\xi^{\nu}_{(1,0),\rho}
\xi^{\mu}_{(1,0),\nu}\right) \nn \\
\fl & + & \frac{\lambda^2\Omega}{2}\left(\xi^{\mu}_{(2,1)}+2\xi^{\nu}_{(1,0)}
\xi^{\mu}_{(1,1),\nu}+\xi^{\nu}_{(0,1)}\xi^{\mu}_{(2,0),\nu}+2\epsilon_2
\xi^{\rho}_{(1,0)}\xi^{\nu}_{(0,1),\rho}\xi^{\mu}_{(1,0),\nu}\right. \nn \\
\fl & + & \left. (\epsilon_1-\epsilon_2)\xi^{\rho}_{(0,1)}
\xi^{\nu}_{(1,0),\rho}\xi^{\mu}_{(1,0),\nu}
+(\epsilon_0-\epsilon_2)\xi^{\rho}_{(1,0)}\xi^{\nu}_{(1,0),\rho}
\xi^{\mu}_{(0,1),\nu}\right) \nn \\
\fl & + & \frac{\lambda\Omega^2}{2}\left(\xi^{\mu}_{(1,2)}+2\xi^{\nu}_{(0,1)}
\xi^{\mu}_{(1,1),\nu}+\xi^{\nu}_{(1,0)}\xi^{\mu}_{(0,2),\nu}+2\epsilon_3
\xi^{\rho}_{(0,1)}\xi^{\nu}_{(1,0),\rho}\xi^{\mu}_{(0,1),\nu}\right. \nn \\
\fl & + & \left. (\epsilon_0-\epsilon_3)\xi^{\rho}_{(1,0)}
\xi^{\nu}_{(0,1),\rho}\xi^{\mu}_{(0,1),\nu}+(\epsilon_1-\epsilon_3)
\xi^{\rho}_{(0,1)}\xi^{\nu}_{(0,1),\rho}\xi^{\mu}_{(1,0),\nu}
\right) \nn \\
\fl & + & \frac{\Omega^3}{6}\left(\xi^{\mu}_{(0,3)}+3\xi^{\nu}_{(0,1)}
\xi^{\mu}_{(0,2),\nu}+\xi^{\rho}_{(0,1)}\xi^{\nu}_{(0,1),\rho}
\xi^{\mu}_{(0,1),\nu}\right) \nn \\
\fl & + & \frac{\lambda^4}{24}\left(\xi^{\mu}_{(4,0)}+
4\xi^{\nu}_{(1,0)}\xi^{\mu}_{(3,0),\nu}
+3\xi^{\nu}_{(2,0)}\xi^{\mu}_{(2,0),\nu}\right. \nn \\
\fl & + & \left. 6\xi^{\rho}_{(1,0)}\xi^{\nu}_{(1,0),\rho}
\xi^{\mu}_{(2,0),\nu}+\xi^{\sigma}_{(1,0)}\xi^{\rho}_{(1,0),\sigma}
\xi^{\nu}_{(1,0),\rho}\xi^{\mu}_{(1,0),\nu})\right) \nn \\
\fl & + & \frac{\lambda^3\Omega}{6}\left(\xi^{\mu}_{(3,1)}+
3\xi^{\nu}_{(1,0)}\xi^{\mu}_{(2,1),\nu}+\xi^{\nu}_{(0,1)}
\xi^{\mu}_{(3,0),\nu}+3\epsilon_4\xi^{\nu}_{(2,0)}\xi^{\mu}_{(1,1),\nu}
\right. \nn \\
\fl & + & 3\epsilon_5\xi^{\nu}_{(1,1)}\xi^{\mu}_{(2,0),\nu}
+3\xi^{\rho}_{(1,0)}\xi^{\nu}_{(1,0),\rho}\xi^{\mu}_{(1,1),\nu}
+3(\epsilon_0\xi^{\rho}_{(1,0)}\xi^{\nu}_{(0,1),\rho}
+\epsilon_1\xi^{\rho}_{(0,1)}\xi^{\nu}_{(1,0),\rho})
\xi^{\mu}_{(2,0),\nu} \nn \\
\fl & + & (\epsilon_1-\epsilon_2-\epsilon_6)\xi^{\sigma}_{(0,1)}
\xi^{\rho}_{(1,0),\sigma}\xi^{\nu}_{(1,0),\rho}\xi^{\mu}_{(1,0),\nu}
+3\epsilon_6\xi^{\sigma}_{(1,0)}\xi^{\rho}_{(0,1),\sigma}
\xi^{\nu}_{(1,0),\rho}\xi^{\mu}_{(1,0),\nu} \nn \\
\fl & + & \left. 3(\epsilon_2-\epsilon_6)\xi^{\sigma}_{(1,0)}
\xi^{\rho}_{(1,0),\sigma}\xi^{\nu}_{(0,1),\rho}\xi^{\mu}_{(1,0),\nu}
+(\epsilon_0-2\epsilon_2+\epsilon_6) \xi^{\sigma}_{(1,0)}
\xi^{\rho}_{(1,0),\sigma}
\xi^{\nu}_{(1,0),\rho}\xi^{\mu}_{(0,1),\nu}\right) \nn \\
\fl & + & \frac{\lambda^2\Omega^2}{4}\left(\xi^{\mu}_{(2,2)}+
2\xi^{\nu}_{(1,0)}\xi^{\mu}_{(1,2),\nu}+2\xi^{\nu}_{(0,1)}
\xi^{\mu}_{(2,1),\nu}+2\xi^{\nu}_{(1,1)}\xi^{\mu}_{(1,1),\nu}\right. \nn \\
\fl & + & \epsilon_7\xi^{\nu}_{(2,0)}\xi^{\mu}_{(0,2),\nu}
+\epsilon_8\xi^{\nu}_{(0,2)}\xi^{\mu}_{(2,0),\nu}
+\xi^{\rho}_{(1,0)}\xi^{\nu}_{(1,0),\rho}\xi^{\mu}_{(0,2),\nu}
+\xi^{\rho}_{(0,1)}\xi^{\nu}_{(0,1),\rho}\xi^{\mu}_{(2,0),\nu} \nn \\
\fl & + & 4(\epsilon_0\xi^{\rho}_{(1,0)}\xi^{\nu}_{(0,1),\rho}
+\epsilon_1\xi^{\rho}_{(0,1)}\xi^{\nu}_{(1,0),\rho})
\xi^{\mu}_{(1,1),\nu} \nn \\
\fl & - & (\epsilon_3+\epsilon_2-\epsilon_1+\epsilon_9)
\xi^{\sigma}_{(0,1)}\xi^{\rho}_{(0,1),\sigma}
\xi^{\nu}_{(1,0),\rho}\xi^{\mu}_{(1,0),\nu} \nn \\
\fl & - & (\epsilon_3+\epsilon_2-\epsilon_0-\epsilon_9)
\xi^{\sigma}_{(1,0)}\xi^{\rho}_{(1,0),\sigma}
\xi^{\nu}_{(0,1),\rho}\xi^{\mu}_{(0,1),\nu} \nn \\
\fl & + & 2(\epsilon_3+\epsilon_2-\epsilon_0\epsilon_1-\epsilon_9)
\xi^{\sigma}_{(1,0)}\xi^{\rho}_{(0,1),\sigma}
\xi^{\nu}_{(1,0),\rho}\xi^{\mu}_{(0,1),\nu} \nn \\
\fl & + & 2(\epsilon_3+\epsilon_2-\epsilon_0\epsilon_1+\epsilon_9)
\xi^{\sigma}_{(0,1)}\xi^{\rho}_{(1,0),\sigma}
\xi^{\nu}_{(0,1),\rho}\xi^{\mu}_{(1,0),\nu} \nn \\
\fl & - & \left. 2(\epsilon_3-\epsilon_0\epsilon_1)
\xi^{\sigma}_{(1,0)}\xi^{\rho}_{(0,1),\sigma}
\xi^{\nu}_{(0,1),\rho}\xi^{\mu}_{(1,0),\nu}
-2(\epsilon_2-\epsilon_0\epsilon_1)
\xi^{\sigma}_{(0,1)}\xi^{\rho}_{(1,0),\sigma}
\xi^{\nu}_{(1,0),\rho}\xi^{\mu}_{(0,1),\nu}\right) \nn \\
\fl & + & \frac{\lambda\Omega^3}{6}\left(\xi^{\mu}_{(1,3)}+
3\xi^{\nu}_{(0,1)}\xi^{\mu}_{(1,2),\nu}+\xi^{\nu}_{(1,0)}
\xi^{\mu}_{(0,3),\nu}+3\epsilon_{10}\xi^{\nu}_{(0,2)}
\xi^{\mu}_{(1,1),\nu}\right. \nn \\
\fl & + & 3\epsilon_{11}\xi^{\nu}_{(1,1)}\xi^{\mu}_{(0,2),\nu}
+3\xi^{\rho}_{(0,1)}\xi^{\nu}_{(0,1),\rho}\xi^{\mu}_{(1,1),\nu}
+3(\epsilon_0\xi^{\rho}_{(1,0)}\xi^{\nu}_{(0,1),\rho}
+\epsilon_1\xi^{\rho}_{(0,1)}\xi^{\nu}_{(1,0),\rho})
\xi^{\mu}_{(0,2),\nu} \nn \\
\fl & + & (\epsilon_0-\epsilon_3-\epsilon_{12})\xi^{\sigma}_{(1,0)}
\xi^{\rho}_{(0,1),\sigma}\xi^{\nu}_{(0,1),\rho}\xi^{\mu}_{(0,1),\nu}
+3\epsilon_{12}\xi^{\sigma}_{(0,1)}\xi^{\rho}_{(1,0),\sigma}
\xi^{\nu}_{(0,1),\rho}\xi^{\mu}_{(0,1),\nu} \nn \\
\fl & + & \left. 3(\epsilon_3-\epsilon_{12})\xi^{\sigma}_{(0,1)}
\xi^{\rho}_{(0,1),\sigma}\xi^{\nu}_{(1,0),\rho}\xi^{\mu}_{(0,1),\nu}
+(\epsilon_1-2\epsilon_3+\epsilon_{12}) \xi^{\sigma}_{(0,1)}
\xi^{\rho}_{(0,1),\sigma}\xi^{\nu}_{(0,1),\rho}\xi^{\mu}_{(1,0),\nu}
\right) \nn \\
\fl & + & \frac{\Omega^4}{24}\left(\xi^{\mu}_{(0,4)}+
4\xi^{\nu}_{(0,1)}\xi^{\mu}_{(0,3),\nu}
+3\xi^{\nu}_{(0,2)}\xi^{\mu}_{(0,2),\nu}\right. \nn \\
\fl & + & \left. 6\xi^{\rho}_{(0,1)}\xi^{\nu}_{(0,1),\rho}
\xi^{\mu}_{(0,2),\nu}+\xi^{\sigma}_{(0,1)}\xi^{\rho}_{(0,1),\sigma}
\xi^{\nu}_{(0,1),\rho}\xi^{\mu}_{(0,1),\nu}\right)  \,,
\label{coordtransf}
\eeq
where the vector fields $\xi^\mu_{(p,q)}$ and their derivatives are
evaluated in $x(p)$.  This expression gives the relation between the
coordinates, in the chart $({\cal U},\mb{X})$, of the two points $p$
and $q$ of $\cM_0$.  
  
On the other hand, the {\it passive} coordinate transformation is
found by inverting~(\ref{coordtransf}):  
\be  
\fl y^{\mu}(q):=x^{\mu}(p)=x^{\mu}(q)-\lambda\xi^{\mu}_{(1,0)}(x(p))-
\Omega\xi^{\mu}_{(0,1)}(x(p))+O^2(\lambda,\Omega) \,,  
\ee  
and then by expanding $x(p)$ around $x(q)$.  We obtain in this way   
an expression of the form   
\be  
\fl y^{\mu}(q)=x^{\mu}(q)-\lambda\xi^{\mu}_{(1,0)}(x(q))-
\Omega\xi^{\mu}_{(0,1)}(x(q))+O^2(\lambda,\Omega)   \,, 
\ee  
which gives the relation between the coordinates of any arbitrary
point $q\in\cM_0$   
in the two charts $({\cal U},\mb{X})$ and $({\cal U}',\mb{Y})$.  Such
a relation is needed to find the transformation of the components 
of a tensor field, by using~(\ref{vectorGTcoords}), as it is usually done 
in textbooks for first order gauge transformations~\cite{dinverno,weinberg}.
However, in order to determine these transformation rules it   
is much simpler to apply directly the
expressions~(\ref{gi10}-\ref{gi04}), computing explicitly the Lie 
derivatives of the tensor field.

%
%

\section{Conclusions}\label{conclusions}

Many astrophysical systems (in particular, oscillating relativistic
rotating stars) can be well described by perturbation
theory depending on two parameters. A well--founded description of 
two--parameter perturbations can  be very useful for such
applications, specially in order to handle properly perturbations at
second order and beyond. For example, one may wish to compare results 
derived in different gauges.

In this paper we have studied the problem of gauge dependence of
non--linear perturbations depending on two parameters, considering
perturbations of arbitrary order in a geometrical perspective, and
generalizing the results of the one--parameter case~\cite{BMMS,SB} to
the case of two parameters.
We  have constructed a geometrical framework in which a {\it gauge choice}
is a two--parameter {\it group} of diffeomorphisms, while a {\it gauge
transformation} is a two--parameter {\it family} of diffeomorphisms.
We have shown that any \mbox{two--parameter} family of diffeomorphisms
can be expanded in terms of Lie derivatives with respect to vectors
$\xi^{\mu}_{(p,q)}$. In terms of this expansion, which can be deduced
order by order, we have derived general expressions for transformations of
coordinates and tensor perturbations, and the conditions for gauge
invariance of tensor perturbations.  We have computed these expressions up
to fourth order in the perturbative expansion, i.e.\ up to terms
$\lambda^k\Omega^{k'}$ with $k+k'=4$.

The way in which the expansion of a two-parameter family of
diffeomorphisms was derived in this paper is order by order,
constructing derivative operators that can be rewritten as Lie
derivatives with respect some vector fields.  The
development of an underlying geometrical structure, analogous to the
knight diffeomorphisms introduced in the one--parameter
case~\cite{BMMS}, would be interesting for two reasons: first,
in order to have a deeper mathematical understanding of the theory,
and second, in order to derive a close formula, valid at all orders,
for gauge transformations and gauge invariance conditions.
The present paper has been devoted to the derivation of the
useful formulae for practical applications. In particular, our 
expressions will be useful to compare results derived in different
gauges, and can form the basis for the construction of a gauge invariant
theory of two-parameter systems in the line of works done for
the one-parameter case like as for 
example~\cite{moncrief,gerlach,bardeen,be,cb}. We leave
the development of a more formal framework for future work.

%
%

\appendix
\section{Proof of the statement (\ref{lieder})}
\label{sebastiano}

The aim of this Appendix is to give a proof of a theorem that
allows us to make the statement contained in
equation~(\ref{lieder}).

~

\noindent {\bf Theorem:} ``Let $\cal L$ be a derivative operator acting
on the set of all the tensor fields defined on a differentiable manifold
$\cal M$ and satisfying the following conditions:
(i) It is linear and satisfies the Leibniz rule; (ii) it is
tensor-type preserving; (iii) it commutes with every
contraction of a tensor field; and (iv) it commutes with the
exterior differentiation d.  Then, there exists a vector field
$\xi$ such that ${\cal L}$ is equivalent to the Lie derivative
operator with respect to $\xi$, that is, $\pounds_\xi$.''

~

First of all, notice that the operators introduced in equations
(\ref{operator10}-\ref{operator04}) satisfy the conditions of 
the theorem.  In particular, properties (iii) and (iv)
follow from the fact that $\Phi^{\ast}$ commutes with contractions
and the exterior derivative (see \cite{thirring}). For more details
on this question see, e.g.,~\cite{kobano}.

The proof of the theorem is as follows: When acting on functions,
$\cal L$ defines a vector field $\xi$ through the relation
\begin{equation}
{\cal L}\,f=:\xi(f), \qquad \forall\; f\in {\cal F}({\cal M})\,,
\label{xi}
\end{equation}
where ${\cal F}({\cal M})$ denotes the algebra of $C^\infty$
functions on ${\cal M}$.  What we want to prove is that on an arbitrary
tensor field $T$,
\begin{equation}
{\cal L}\,T=\pounds_\xi\,T\,.\label{29}
\end{equation}
Clearly (\ref{29}) holds for an arbitrary tensor field T iff it holds
for an arbitrary vector field V. For the latter equation (\ref{29}) is
equivalent to the following expression
\beq
{\cal L}\,V=[\xi,V]\,.
\eeq
Applying this to any function $f$ we obtain
\be
({\cal L}\,V)(f)=\xi[V(f)]-V[\xi(f)]\,,\qquad \forall\; f\in {\cal F}
({\cal M})\,. \label{vecf}
\ee
Therefore, to prove (\ref{29}) is equivalent to prove (\ref{vecf}).
To this end, let us consider the action of the operator $\cal L$
on the function $V(f)$.  Using~(\ref{xi}) we have
\begin{equation}
{\cal L}[V(f)]=\xi[V(f)] \,. \label{1}
\end{equation}
On the other hand, using the properties (i)-(iv) of ${\cal L}$ we
have
\beq
\fl {\cal L}[V(f)] & = & {\cal L}({\rm d}f(V))={\cal L}[{\cal C}
({\rm d}f\otimes V)]=
{\cal C}[{\rm d}({\cal L}f)\otimes V + {\rm d}f\otimes {\cal L}V] \nn \\
\fl & = & {\rm d}({\cal L}f)V + {\rm d}f({\cal L}V) =
V({\cal L}f) + ({\cal L}V)(f)\,.
\eeq
Then, this in combination with (\ref{1}), and using (\ref{xi}),
leads to equation~(\ref{vecf}), which is what we wanted to prove.

%
%

\section*{Acknowledgments}
We thank Sebastiano Sonego for his help in the proof of equation
(\ref{lieder}). We thank the anonymous referees for suggesting changes
that have improved the manuscript.
MB and LG thank Valeria Ferrari for useful comments and remarks.
MB thanks the Dipartimento di Fisica ``G. Marconi'' (Universit\`a di Roma
``La Sapienza'') and LG the Institute of Cosmology and Gravitation
(University of Portsmouth) for hospitality during some stages in the
realization of this work.
CFS is supported by the EPSRC. This work has been supported by the EU
programme `Improving the Human Research Potential and the Socio--Economic
Knowledge Base' (Research Training Network Contract HPRN--CT--2000--00137).

%
%


\end{document}